\documentclass[draftcls,12pt,onecolumn,twoside]{IEEEtran}
\usepackage[sort]{cite}
\usepackage[T1]{fontenc}
\usepackage{graphicx}  
\usepackage{caption}
\usepackage{subcaption}
\usepackage[table]{xcolor}

\usepackage{amssymb}
\usepackage{amsmath}
\usepackage{paralist}

\usepackage{url}
\usepackage{booktabs}
\usepackage{multirow}
\usepackage{pgfplots}

\usepackage{float}
\usepackage{bbm}
\floatstyle{ruled}
\newfloat{model}{thp}{lom}
\floatname{model}{Model}

\usepgfplotslibrary{external} 
\usetikzlibrary{pgfplots.external} 
\usepackage{amssymb}
\usepackage{amsfonts}
\usepackage{mathrsfs}
\usepackage{xspace}
\usepackage{bm}
\usepackage{upgreek}

\newcommand{\safemath}[2]{\newcommand{#1}{\ensuremath{#2}\xspace}}

\safemath{\bma}{\mathbf{a}}
\safemath{\bmb}{\mathbf{b}}
\safemath{\bmc}{\mathbf{c}}
\safemath{\bmd}{\mathbf{d}}
\safemath{\bme}{\mathbf{e}}
\safemath{\bmf}{\mathbf{f}}
\safemath{\bmg}{\mathbf{g}}
\safemath{\bmh}{\mathbf{h}}
\safemath{\bmi}{\mathbf{i}}
\safemath{\bmj}{\mathbf{j}}
\safemath{\bmk}{\mathbf{k}}
\safemath{\bml}{\mathbf{l}}
\safemath{\bmm}{\mathbf{m}}
\safemath{\bmn}{\mathbf{n}}
\safemath{\bmo}{\mathbf{o}}
\safemath{\bmp}{\mathbf{p}}
\safemath{\bmq}{\mathbf{q}}
\safemath{\bmr}{\mathbf{r}}
\safemath{\bms}{\mathbf{s}}
\safemath{\bmt}{\mathbf{t}}
\safemath{\bmu}{\mathbf{u}}
\safemath{\bmv}{\mathbf{v}}
\safemath{\bmw}{\mathbf{w}}
\safemath{\bmx}{\mathbf{x}}
\safemath{\bmy}{\mathbf{y}}
\safemath{\bmz}{\mathbf{z}}
\safemath{\bmzero}{\mathbf{0}}
\safemath{\bmone}{\mathbf{1}}

\bmdefine{\biad}{a}
\bmdefine{\bibd}{b}
\bmdefine{\bicd}{c}
\bmdefine{\bidd}{d}
\bmdefine{\bied}{e}
\bmdefine{\bifd}{f}
\bmdefine{\bigd}{g}
\bmdefine{\bihd}{h}
\bmdefine{\biid}{i}
\bmdefine{\bijd}{j}
\bmdefine{\bikd}{k}
\bmdefine{\bild}{l}
\bmdefine{\bimd}{m}
\bmdefine{\bind}{n}
\bmdefine{\biod}{o}
\bmdefine{\bipd}{p}
\bmdefine{\biqd}{q}
\bmdefine{\bird}{r}
\bmdefine{\bisd}{s}
\bmdefine{\bitd}{t}
\bmdefine{\biud}{u}
\bmdefine{\bivd}{v}
\bmdefine{\biwd}{w}
\bmdefine{\bixd}{x}
\bmdefine{\biyd}{y}
\bmdefine{\bizd}{z}

\bmdefine{\bixid}{\xi}
\bmdefine{\bilambdad}{\lambda}
\bmdefine{\bimud}{\mu}
\bmdefine{\bithetad}{\theta}
\bmdefine{\biphid}{\phi}
\bmdefine{\bideltad}{\delta}

\safemath{\bmia}{\biad}
\safemath{\bmib}{\bibd}
\safemath{\bmic}{\bicd}
\safemath{\bmid}{\bidd}
\safemath{\bmie}{\bied}
\safemath{\bmif}{\bifd}
\safemath{\bmig}{\bigd}
\safemath{\bmih}{\bihd}
\safemath{\bmii}{\biid}
\safemath{\bmij}{\bijd}
\safemath{\bmik}{\bikd}
\safemath{\bmil}{\bild}
\safemath{\bmim}{\bimd}
\safemath{\bmin}{\bind}
\safemath{\bmio}{\biod}
\safemath{\bmip}{\bipd}
\safemath{\bmiq}{\biqd}
\safemath{\bmir}{\bird}
\safemath{\bmis}{\bisd}
\safemath{\bmit}{\bitd}
\safemath{\bmiu}{\biud}
\safemath{\bmiv}{\bivd}
\safemath{\bmiw}{\biwd}
\safemath{\bmix}{\bixd}
\safemath{\bmiy}{\biyd}
\safemath{\bmiz}{\bizd}

\safemath{\bmxi}{\bixid}
\safemath{\bmlambda}{\bilambdad}
\safemath{\bmmu}{\bimud}
\safemath{\bmtheta}{\bithetad}
\safemath{\bmphi}{\biphid}
\safemath{\bmdelta}{\bideltad}

\safemath{\bA}{\mathbf{A}}
\safemath{\bB}{\mathbf{B}}
\safemath{\bC}{\mathbf{C}}
\safemath{\bD}{\mathbf{D}}
\safemath{\bE}{\mathbf{E}}
\safemath{\bF}{\mathbf{F}}
\safemath{\bG}{\mathbf{G}}
\safemath{\bH}{\mathbf{H}}
\safemath{\bI}{\mathbf{I}}
\safemath{\bJ}{\mathbf{J}}
\safemath{\bK}{\mathbf{K}}
\safemath{\bL}{\mathbf{L}}
\safemath{\bM}{\mathbf{M}}
\safemath{\bN}{\mathbf{N}}
\safemath{\bO}{\mathbf{O}}
\safemath{\bP}{\mathbf{P}}
\safemath{\bQ}{\mathbf{Q}}
\safemath{\bR}{\mathbf{R}}
\safemath{\bS}{\mathbf{S}}
\safemath{\bT}{\mathbf{T}}
\safemath{\bU}{\mathbf{U}}
\safemath{\bV}{\mathbf{V}}
\safemath{\bW}{\mathbf{W}}
\safemath{\bX}{\mathbf{X}}
\safemath{\bY}{\mathbf{Y}}
\safemath{\bZ}{\mathbf{Z}}

\safemath{\bZero}{\mathbf{0}}
\safemath{\bOne}{\mathbf{1}}
\safemath{\bDelta}{\mathbf{\Delta}}
\safemath{\bLambda}{\mathbf{\UpLambda}}
\safemath{\bPhi}{\mathbf{\Upphi}}
\safemath{\bSigma}{\mathbf{\Upsigma}}
\safemath{\bOmega}{\mathbf{\Upomega}}
\safemath{\bTheta}{\mathbf{\Uptheta}}

\bmdefine{\biAd}{A}
\bmdefine{\biBd}{B}
\bmdefine{\biCd}{C}
\bmdefine{\biDd}{D}
\bmdefine{\biEd}{E}
\bmdefine{\biFd}{F}
\bmdefine{\biGd}{G}
\bmdefine{\biHd}{H}
\bmdefine{\biId}{I}
\bmdefine{\biJd}{J}
\bmdefine{\biKd}{K}
\bmdefine{\biLd}{L}
\bmdefine{\biMd}{M}
\bmdefine{\biOd}{N}
\bmdefine{\biPd}{O}
\bmdefine{\biQd}{P}
\bmdefine{\biRd}{R}
\bmdefine{\biSd}{S}
\bmdefine{\biTd}{T}
\bmdefine{\biUd}{U}
\bmdefine{\biVd}{V}
\bmdefine{\biWd}{W}
\bmdefine{\biXd}{X}
\bmdefine{\biYd}{Y}
\bmdefine{\biZd}{Z}

\bmdefine{\biDelta}{\Delta}
\bmdefine{\biLambda}{\Lambda}
\bmdefine{\biPhi}{\Phi}
\bmdefine{\biSigma}{\Sigma}
\bmdefine{\biOmega}{\Omega}
\bmdefine{\biTheta}{\Theta}

\safemath{\bimA}{\biAd}
\safemath{\bimB}{\biBd}
\safemath{\bimC}{\biCd}
\safemath{\bimD}{\biDd}
\safemath{\bimE}{\biEd}
\safemath{\bimF}{\biFd}
\safemath{\bimG}{\biGd}
\safemath{\bimH}{\biHd}
\safemath{\bimI}{\biId}
\safemath{\bimJ}{\biJd}
\safemath{\bimK}{\biKd}
\safemath{\bimL}{\biLd}
\safemath{\bimM}{\biMd}
\safemath{\bimN}{\biNd}
\safemath{\bimO}{\biOd}
\safemath{\bimP}{\biPd}
\safemath{\bimQ}{\biQd}
\safemath{\bimR}{\biRd}
\safemath{\bimS}{\biSd}
\safemath{\bimT}{\biTd}
\safemath{\bimU}{\biUd}
\safemath{\bimV}{\biVd}
\safemath{\bimW}{\biWd}
\safemath{\bimX}{\biXd}
\safemath{\bimY}{\biYd}
\safemath{\bimZ}{\biZd}

\safemath{\bimDelta}{\biDelta}
\safemath{\bimLambda}{\biLambda}
\safemath{\bimPhi}{\biPhi}
\safemath{\bimSigma}{\biSigma}
\safemath{\bimOmega}{\biOmega}
\safemath{\bimTheta}{\biTheta}

\safemath{\setA}{\mathcal{A}}
\safemath{\setB}{\mathcal{B}}
\safemath{\setC}{\mathcal{C}}
\safemath{\setD}{\mathcal{D}}
\safemath{\setE}{\mathcal{E}}
\safemath{\setF}{\mathcal{F}}
\safemath{\setG}{\mathcal{G}}
\safemath{\setH}{\mathcal{H}}
\safemath{\setI}{\mathcal{I}}
\safemath{\setJ}{\mathcal{J}}
\safemath{\setK}{\mathcal{K}}
\safemath{\setL}{\mathcal{L}}
\safemath{\setM}{\mathcal{M}}
\safemath{\setN}{\mathcal{N}}
\safemath{\setO}{\mathcal{O}}
\safemath{\setP}{\mathcal{P}}
\safemath{\setQ}{\mathcal{Q}}
\safemath{\setR}{\mathcal{R}}
\safemath{\setS}{\mathcal{S}}
\safemath{\setT}{\mathcal{T}}
\safemath{\setU}{\mathcal{U}}
\safemath{\setV}{\mathcal{V}}
\safemath{\setW}{\mathcal{W}}
\safemath{\setX}{\mathcal{X}}
\safemath{\setY}{\mathcal{Y}}
\safemath{\setZ}{\mathcal{Z}}
\safemath{\emptySet}{\varnothing}

\safemath{\colA}{\mathscr{A}}
\safemath{\colB}{\mathscr{B}}
\safemath{\colC}{\mathscr{C}}
\safemath{\colD}{\mathscr{D}}
\safemath{\colE}{\mathscr{E}}
\safemath{\colF}{\mathscr{F}}
\safemath{\colG}{\mathscr{G}}
\safemath{\colH}{\mathscr{H}}
\safemath{\colI}{\mathscr{I}}
\safemath{\colJ}{\mathscr{J}}
\safemath{\colK}{\mathscr{K}}
\safemath{\colL}{\mathscr{L}}
\safemath{\colM}{\mathscr{M}}
\safemath{\colN}{\mathscr{N}}
\safemath{\colO}{\mathscr{O}}
\safemath{\colP}{\mathscr{P}}
\safemath{\colQ}{\mathscr{Q}}
\safemath{\colR}{\mathscr{R}}
\safemath{\colS}{\mathscr{S}}
\safemath{\colT}{\mathscr{T}}
\safemath{\colU}{\mathscr{U}}
\safemath{\colV}{\mathscr{V}}
\safemath{\colW}{\mathscr{W}}
\safemath{\colX}{\mathscr{X}}
\safemath{\colY}{\mathscr{Y}}
\safemath{\colZ}{\mathscr{Z}}

\safemath{\opA}{\mathbb{A}}
\safemath{\opB}{\mathbb{B}}
\safemath{\opC}{\mathbb{C}}
\safemath{\opD}{\mathbb{D}}
\safemath{\opE}{\mathbb{E}}
\safemath{\opF}{\mathbb{F}}
\safemath{\opG}{\mathbb{G}}
\safemath{\opH}{\mathbb{H}}
\safemath{\opI}{\mathbb{I}}
\safemath{\opJ}{\mathbb{J}}
\safemath{\opK}{\mathbb{K}}
\safemath{\opL}{\mathbb{L}}
\safemath{\opM}{\mathbb{M}}
\safemath{\opN}{\mathbb{N}}
\safemath{\opO}{\mathbb{O}}
\safemath{\opP}{\mathbb{P}}
\safemath{\opQ}{\mathbb{Q}}
\safemath{\opR}{\mathbb{R}}
\safemath{\opS}{\mathbb{S}}
\safemath{\opT}{\mathbb{T}}
\safemath{\opU}{\mathbb{U}}
\safemath{\opV}{\mathbb{V}}
\safemath{\opW}{\mathbb{W}}
\safemath{\opX}{\mathbb{X}}
\safemath{\opY}{\mathbb{Y}}
\safemath{\opZ}{\mathbb{Z}}
\safemath{\opZero}{\mathbb{O}}
\safemath{\identityop}{\opI}

\safemath{\veca}{\bma}
\safemath{\vecb}{\bmb}
\safemath{\vecc}{\bmc}
\safemath{\vecd}{\bmd}
\safemath{\vece}{\bme}
\safemath{\vecf}{\bmf}
\safemath{\vecg}{\bmg}
\safemath{\vech}{\bmh}
\safemath{\veci}{\bmi}
\safemath{\vecj}{\bmj}
\safemath{\veck}{\bmk}
\safemath{\vecl}{\bml}
\safemath{\vecm}{\bmm}
\safemath{\vecn}{\bmn}
\safemath{\veco}{\bmo}
\safemath{\vecp}{\bmp}
\safemath{\vecq}{\bmq}
\safemath{\vecr}{\bmr}
\safemath{\vecs}{\bms}
\safemath{\vect}{\bmt}
\safemath{\vecu}{\bmu}
\safemath{\vecv}{\bmv}
\safemath{\vecw}{\bmw}
\safemath{\vecx}{\bmx}
\safemath{\vecy}{\bmy}
\safemath{\vecz}{\bmz}

\safemath{\veczero}{\bmzero}
\safemath{\vecone}{\bmone}
\safemath{\vecxi}{\bmxi}
\safemath{\veclambda}{\bmlambda}
\safemath{\vecmu}{\bmmu}
\safemath{\vectheta}{\bmtheta}
\safemath{\vecphi}{\bmphi}
\safemath{\vecdelta}{\bmdelta}

\safemath{\matA}{\bA}
\safemath{\matB}{\bB}
\safemath{\matC}{\bC}
\safemath{\matD}{\bD}
\safemath{\matE}{\bE}
\safemath{\matF}{\bF}
\safemath{\matG}{\bG}
\safemath{\matH}{\bH}
\safemath{\matI}{\bI}
\safemath{\matJ}{\bJ}
\safemath{\matK}{\bK}
\safemath{\matL}{\bL}
\safemath{\matM}{\bM}
\safemath{\matN}{\bN}
\safemath{\matO}{\bO}
\safemath{\matP}{\bP}
\safemath{\matQ}{\bQ}
\safemath{\matR}{\bR}
\safemath{\matS}{\bS}
\safemath{\matT}{\bT}
\safemath{\matU}{\bU}
\safemath{\matV}{\bV}
\safemath{\matW}{\bW}
\safemath{\matX}{\bX}
\safemath{\matY}{\bY}
\safemath{\matZ}{\bZ}
\safemath{\matzero}{\bmzero}

\safemath{\matDelta}{\bDelta}
\safemath{\matLambda}{\bLambda}
\safemath{\matPhi}{\bPhi}
\safemath{\matSigma}{\bSigma}
\safemath{\matOmega}{\bOmega}
\safemath{\matTheta}{\bTheta}

\safemath{\matidentity}{\matI}
\safemath{\matone}{\matO}

\safemath{\rnda}{A}
\safemath{\rndb}{B}
\safemath{\rndc}{C}
\safemath{\rndd}{D}
\safemath{\rnde}{E}
\safemath{\rndf}{F}
\safemath{\rndg}{G}
\safemath{\rndh}{H}
\safemath{\rndi}{I}
\safemath{\rndj}{J}
\safemath{\rndk}{K}
\safemath{\rndl}{L}
\safemath{\rndm}{M}
\safemath{\rndn}{N}
\safemath{\rndo}{O}
\safemath{\rndp}{P}
\safemath{\rndq}{Q}
\safemath{\rndr}{R}
\safemath{\rnds}{S}
\safemath{\rndt}{T}
\safemath{\rndu}{U}
\safemath{\rndv}{V}
\safemath{\rndw}{W}
\safemath{\rndx}{X}
\safemath{\rndy}{Y}
\safemath{\rndz}{Z}

\safemath{\rveca}{\bimA}
\safemath{\rvecb}{\bimB}
\safemath{\rvecc}{\bimC}
\safemath{\rvecd}{\bimD}
\safemath{\rvece}{\bimE}
\safemath{\rvecf}{\bimF}
\safemath{\rvecg}{\bimG}
\safemath{\rvech}{\bimH}
\safemath{\rveci}{\bimI}
\safemath{\rvecj}{\bimJ}
\safemath{\rveck}{\bimK}
\safemath{\rvecl}{\bimL}
\safemath{\rvecm}{\bimM}
\safemath{\rvecn}{\bimN}
\safemath{\rveco}{\bomO}
\safemath{\rvecp}{\bimP}
\safemath{\rvecq}{\bimQ}
\safemath{\rvecr}{\bimR}
\safemath{\rvecs}{\bimS}
\safemath{\rvect}{\bimT}
\safemath{\rvecu}{\bimU}
\safemath{\rvecv}{\bimV}
\safemath{\rvecw}{\bimW}
\safemath{\rvecx}{\bimX}
\safemath{\rvecy}{\bimY}
\safemath{\rvecz}{\bimZ}

\safemath{\rvecxi}{\bmxi}
\safemath{\rveclambda}{\bmlambda}
\safemath{\rvecmu}{\bmmu}
\safemath{\rvectheta}{\bmtheta}
\safemath{\rvecphi}{\bmphi}

\safemath{\rmatA}{\bimA}
\safemath{\rmatB}{\bimB}
\safemath{\rmatC}{\bimC}
\safemath{\rmatD}{\bimD}
\safemath{\rmatE}{\bimE}
\safemath{\rmatF}{\bimF}
\safemath{\rmatG}{\bimG}
\safemath{\rmatH}{\bimH}
\safemath{\rmatI}{\bimI}
\safemath{\rmatJ}{\bimJ}
\safemath{\rmatK}{\bimK}
\safemath{\rmatL}{\bimL}
\safemath{\rmatM}{\bimM}
\safemath{\rmatN}{\bimN}
\safemath{\rmatO}{\bimO}
\safemath{\rmatP}{\bimP}
\safemath{\rmatQ}{\bimQ}
\safemath{\rmatR}{\bimR}
\safemath{\rmatS}{\bimS}
\safemath{\rmatT}{\bimT}
\safemath{\rmatU}{\bimU}
\safemath{\rmatV}{\bimV}
\safemath{\rmatW}{\bimW}
\safemath{\rmatX}{\bimX}
\safemath{\rmatY}{\bimY}
\safemath{\rmatZ}{\bimZ}

\safemath{\rmatDelta}{\bimDelta}
\safemath{\rmatLambda}{\bimLambda}
\safemath{\rmatPhi}{\bimPhi}
\safemath{\rmatSigma}{\bimSigma}
\safemath{\rmatOmega}{\bimOmega}
\safemath{\rmatTheta}{\bimTheta}

\usepackage{amssymb}
\usepackage{amsfonts}
\usepackage{mathrsfs}
\usepackage{xspace}
\usepackage{bm}
\usepackage{fancyref}
\usepackage{textcomp}

\usepackage{multirow}
\usepackage{stmaryrd}

\usepackage{functan}

\newenvironment{textbmatrix}{	\setlength{\arraycolsep}{2.5pt}%
								\big[\begin{matrix}}{\end{matrix}\big]%
								\raisebox{0.08ex}{\vphantom{M}}}
				
\Macro{l2l2}{2,2}
\Macro{l1l2}{1,2}
\Macro{l1l1}{1,1}
\Macro{li}{\infty}		
\Macro{l2}{2}		
\Macro{l1}{1}		
\Macro{l0}{0}				

\def\be{\begin{equation}}
\def\ee{\end{equation}}
\def\een{\nonumber \end{equation}}
\def\mat{\begin{bmatrix}}
\def\emat{\end{bmatrix}}
\def\btm{\begin{textbmatrix}}
\def\etm{\end{textbmatrix}}

\def\ba#1\ea{\begin{align}#1\end{align}}
\def\bas#1\eas{\begin{align*}#1\end{align*}}
\def\bs#1\es{\begin{split}#1\end{split}} 
\def\bg#1\eg{\begin{gather}#1\end{gather}}
\def\bml#1\eml{\begin{multline}#1\end{multline}}
\def\bi#1\ei{\begin{itemize}#1\end{itemize}}

\newcommand{\lefto}{\mathopen{}\left}

\DeclareMathOperator*{\mini}{\textrm{minimize}}
\DeclareMathOperator*{\st}{\textrm{subject to}}

\DeclareMathOperator{\rank}{rank}			
\DeclareMathOperator{\adj}{adj}				
\DeclareMathOperator{\sign}{sign}			

\newcommand{\abs}[1]{\lefto\lvert#1\right\rvert}		

\newcommand{\herm}[1]{\ensuremath{#1^{H}}} 	
\newcommand{\pinv}[1]{\ensuremath{#1^{\dagger}}} 	

\safemath{\dirac}{\delta}					
\safemath{\krond}{\dirac}					

\safemath{\upto}{\uparrow}
\safemath{\downto}{\downarrow}
\safemath{\iu}{j}							
\safemath{\ev}{\lambda}						
\safemath{\hilseqspace}{l^{2}}				
\newcommand{\banachfunspace}[1]{\setL^{#1}}	
\safemath{\hilfunspace}{\banachfunspace{2}}	

\safemath{\SNR}{\text{\sc snr}} 				
\safemath{\No}{N_0}							
\safemath{\Es}{E_s}							
\safemath{\Eb}{E_b}							
\safemath{\EbNo}{\frac{\Eb}{\No}}
\safemath{\EsNo}{\frac{\Es}{\No}}

\DeclareMathOperator{\CHop}{\ensuremath{\opH}} 
\safemath{\tvir}{\rndh_{\CHop}}				
\safemath{\tvtf}{\rndl_{\CHop}}				
\safemath{\spf}{\rnds_{\CHop}}				
\safemath{\bff}{H_{\CHop}}					

\safemath{\ircf}{r_{h}}						
\safemath{\tftvcf}{r_{s}}					
\safemath{\tfcf}{r_{l}}						
\safemath{\bfcf}{r_{H}}						

\safemath{\tcorr}{c_h}						
\safemath{\scf}{c_{s}}						
\safemath{\tfcorr}{c_{l}}					
\safemath{\fcorr}{c_{H}}						

\safemath{\mi}{I}							
\safemath{\capacity}{C}						

\safemath{\normal}{\mathcal{N}}			
\safemath{\jpg}{\mathcal{CN}}			
\safemath{\mchain}{\leftrightarrow}		

\safemath{\dB}{\,\mathrm{dB}}
\safemath{\dBm}{\,\mathrm{dBm}}
\safemath{\Hz}{\,\mathrm{Hz}}
\safemath{\kHz}{\,\mathrm{kHz}}
\safemath{\MHz}{\,\mathrm{MHz}}
\safemath{\GHz}{\,\mathrm{GHz}}
\safemath{\s}{\,\mathrm{s}}
\safemath{\ms}{\,\mathrm{ms}}
\safemath{\mus}{\,\mathrm{\text{\textmu}s}}
\safemath{\ns}{\,\mathrm{ns}}
\safemath{\ps}{\,\mathrm{ps}}
\safemath{\meter}{\,\mathrm{m}}
\safemath{\mm}{\,\mathrm{mm}}
\safemath{\cm}{\,\mathrm{cm}}
\safemath{\W}{\,\mathrm{W}}
\safemath{\mW}{\, \mathrm{mW}}
\safemath{\J}{\,\mathrm{J}}
\safemath{\K}{\,\mathrm{K}}
\safemath{\bit}{\,\mathrm{bit}}
\safemath{\nat}{\,\mathrm{nat}}

\safemath{\define}{=}			

\safemath{\equivalent}{\sim}
\safemath{\distas}{\sim}					
\safemath{\sdiff}{\Delta}				

\safemath{\reals}{\mathbb{R}}
\safemath{\positivereals}{\reals_{+}}
\safemath{\integers}{\mathbb{Z}}
\safemath{\posint}{\integers_{+}}
\safemath{\naturals}{\mathbb{N}}
\safemath{\posnaturals}{\naturals_{+}}
\safemath{\complexset}{\mathbb{C}}
\safemath{\rationals}{\mathbb{Q}}

\newcommand*{\fancyrefapplabelprefix}{app}		
\newcommand*{\fancyrefthmlabelprefix}{thm}		
\newcommand*{\fancyreflemlabelprefix}{lem}		
\newcommand*{\fancyrefcorlabelprefix}{cor}		
\newcommand*{\fancyrefdeflabelprefix}{def}		
\newcommand*{\fancyrefproplabelprefix}{prop}		
\newcommand*{\fancyrefexmpllabelprefix}{exmpl}
\frefformat{vario}{\fancyrefseclabelprefix}{Section~#1}
\frefformat{vario}{\fancyrefthmlabelprefix}{Theorem~#1}
\frefformat{vario}{\fancyreflemlabelprefix}{Lemma~#1}
\frefformat{vario}{\fancyrefcorlabelprefix}{Corollary~#1}
\frefformat{vario}{\fancyrefdeflabelprefix}{Definition~#1}
\frefformat{vario}{\fancyreffiglabelprefix}{Figure~#1}
\frefformat{vario}{\fancyrefeqlabelprefix}{(#1)}
\frefformat{vario}{\fancyrefproplabelprefix}{Property~#1}
\frefformat{vario}{\fancyrefexmpllabelprefix}{Example~#1}
\frefformat{vario}{\fancyrefapplabelprefix}{Appendix~#1}
\frefformat{vario}{\fancyreftablabelprefix}{Table~#1}

 \newtheorem{thm}{Theorem}

 \newtheorem{lem}[thm]{Lemma}
 
\renewcommand{\le}{\leqslant}
\renewcommand{\ge}{\geqslant}

\safemath{\cplxi}{\imath}
\safemath{\cplxj}{\jmath}

\renewcommand{\adj}{{H}}
\newcommand{\trans}{{T}}

\newcommand{\prob}[1]{\mathbb{P}\lefto\{ #1 \right\}}
\newcommand{\expected}[1]{\mathbb{E}\lefto[ #1 \right]}
\newcommand{\moment}[2]{\mathbb{E}^{#1}\lefto[ #2 \right]}
\newcommand{\momentx}[3]{\mathbb{E}^{#1}_{#2}\lefto[ #3 \right]}

\newcommand{\projse}{\matP_{\sigsup,\errsup}}
\newcommand{\projset}{\tilde{\matP}_{\sigsup,\errsup}}

\newcommand{\identity}{\matI}

\newcommand{\rest}{\matR}

\newcommand{\eps}{\varepsilon}

\newcommand{\Gersgorin}{Ger\v{s}gorin\xspace}

\newcommand{\XkaEka}{1a\xspace}
\newcommand{\XuaEka}{2a\xspace}
\newcommand{\XkaEua}{2a\xspace}
\newcommand{\XuaEua}{3a\xspace}

\newcommand{\XkrEka}{1b\xspace}
\newcommand{\XurEka}{2b\xspace}
\newcommand{\XkrEua}{2c\xspace}
\newcommand{\XurEua}{3b\xspace}

\newcommand{\XkaEkr}{1b\xspace}
\newcommand{\XuaEkr}{2c\xspace}
\newcommand{\XkaEur}{2b\xspace}
\newcommand{\XuaEur}{3b\xspace}

\newcommand{\XkrEkr}{1c\xspace}
\newcommand{\XurEkr}{2d\xspace}
\newcommand{\XkrEur}{2d\xspace}
\newcommand{\XurEur}{3c\xspace}

\newcommand{\isonb}{\mathbbm{1}[{\coh\neq 0}]}
\newcommand{\isonba}{\mathbbm{1}[{\coha\neq 0}]}
\newcommand{\isonbb}{\mathbbm{1}[{\cohb\neq 0}]}

\safemath{\dict}{\matD}
\safemath{\dictt}{\tilde{\dict}}
\safemath{\dcol}{\vecd}
\safemath{\dicta}{\matA}
\safemath{\dictaa}{\tilde{\dicta}}
\safemath{\dictbb}{\tilde{\dictb}}
\newcommand{\dictset}{\tilde{\dict}_{\sigsup,\errsup}}
\safemath{\dictat}{\tilde{\dicta}}
\safemath{\cola}{\veca}
\safemath{\dictb}{\matB}
\safemath{\colb}{\vecb}

\safemath{\sig}{\vecx}
\safemath{\obs}{\vecz}
\safemath{\err}{\vece}
\safemath{\noise}{\vecn}
\safemath{\sigsup}{\setX}
\safemath{\errsup}{\setE}
\safemath{\nsig}{n_x}
\safemath{\nerr}{n_e}

\newcommand{\dictse}{\dict_{\sigsup,\errsup}}

\newcommand{\ssig}{\vecs}
\newcommand{\ssigse}{\vecs_{\sigsup,\errsup}}

\newcommand{\setAss}{\setA^\errsup_{\sigsup,\sigsup'}}

\safemath{\Pzero}{\textrm{P0}}	
\safemath{\Pone}{\textrm{BP}}

\safemath{\PzeroE}{(\Pzero,\setE)}
\safemath{\PoneE}{(\Pone,\setE)}
\safemath{\PzeroX}{(\Pzero,\setX)}
\safemath{\PoneX}{(\Pone,\setX)}

\safemath{\PzeroEs}{(\Pzero^\star,\setE)}
\safemath{\PoneEs}{(\Pone^\star,\setE)}
\safemath{\PzeroXs}{(\Pzero^\star,\setX)}
\safemath{\PoneXs}{(\Pone^\star,\setX)}

\safemath{\BP}{(\textrm{BP})}
\safemath{\BPDN}{(\textrm{BPDN})}
\safemath{\PZ}{(\textrm{P0})}
\safemath{\PO}{(\textrm{P1})}
\safemath{\PZC}{(\textrm{P0}^\star)}
\safemath{\BPC}{(\textrm{BP}^\star)}

\safemath{\dima}{{n_a}}
\safemath{\dimb}{{n_b}}
\safemath{\dimm}{m}
\safemath{\dimn}{n}
\safemath{\sparsity}{s}
\safemath{\support}{\setS}

\safemath{\pzeromodel}{\mathcal{M}(\Pzero)}
\safemath{\ponemodel}{\mathcal{M}(\Pone)}

\newcommand{\event}{\mathfrak{E}}
\newcommand{\fr}{\mathfrak{R}}

\safemath{\coh}{\mu}  
\safemath{\coha}{\coh_a}
\safemath{\cohb}{\coh_b}
\safemath{\cohd}{\coh_d}
\safemath{\mcoh}{\coh_m}
\safemath{\cum}{\nu}  

\safemath{\const}{\mathrm{C}}

\newcommand{\smin}{\sigma_{\min}}
\newcommand{\smax}{\sigma_{\max}}

\newcommand{\range}{\mathcal{R}}
\DeclareMathOperator{\supp}{supp}

\tikzset{every mark/.append style=solid}

\pgfplotscreateplotcyclelist{bw3}{%
   {black,thick},
   {black,thick,dashed}, 
   {black,thick,densely dotted},  
    }

\pgfplotscreateplotcyclelist{bw32}{%
   {black,thick},
   {black,thick,dashed}, 
   {black,thick,densely dotted},  
   {black,thick, mark=o},
   {black,thick,dashed, mark=o}, 
   {black,thick,densely dotted, mark=o},  
    }

\pgfplotscreateplotcyclelist{colour3}{%
   {blue!50!black,very thick,densely dashed},           
   {black,very thick},
   {teal!50!black,very thick,densely dotted}, 
   {blue,very thick,densely dashed},           
   {black!75!white,very thick},
   {teal,very thick,densely dotted}, 
   {blue!50!white,very thick,densely dashed},           
   {black!50!white,very thick},
   {teal!75!white,very thick,densely dotted},
   }

\pgfplotscreateplotcyclelist{colour2}{%
   {blue!50!black, thick, mark = o},           
   {black, thick},
   {blue, thick, mark = triangle},           
   {black!75!white, thick},
   {blue!50!white, thick, mark=square},           
   {black!50!white, thick},
   }

\newcommand{\newres}[1]{\cellcolor[gray]{0.8}\textcolor{black}{\textbf{#1}}}
\newcommand{\newresb}[1]{\cellcolor[gray]{0.8}\textcolor{black}{{#1}}}

\newcommand{\improvres}[1]{\cellcolor[gray]{0.9}\textcolor{black}{\textbf{#1}}}
\newcommand{\improvresb}[1]{\cellcolor[gray]{0.9}\textcolor{black}{{#1}}}

\newcommand{\oldres}[1]{\cellcolor[gray]{1}\textcolor{black}{#1} }

\usepackage[normalem]{ulem}	
\newcommand{\rev}[1]{\textcolor{red}{#1}}
\renewcommand{\rev}[1]{#1}

\pgfplotsset{
    tick label style={font=\footnotesize},
    label style={font=\small},
    legend style={font=\footnotesize}
}

\begin{document}

\title{Probabilistic Recovery Guarantees \\ for Sparsely Corrupted Signals}

\author{Graeme Pope,~\IEEEmembership{Student Member,~IEEE}, Annina Bracher, and Christoph~Studer,~\IEEEmembership{Member,~IEEE}

\thanks{\rev{The material in this paper was presented in part at the IEEE Information Theory Workshop (ITW), September, Lausanne, Switzerland, 2012\cite{Bracher2012}}.}
\thanks{G.~Pope and A.~Bracher are with the Dept.~of Information Technology and Electrical Engineering, ETH Zurich, 8092 Zurich, Switzerland (e-mail:~\mbox{gpope@nari.ee.ethz.ch}, \mbox{brachera@student.ethz.ch}).}
\thanks{ C.~Studer is with the  Dept.~of Electrical and Computer Engineering, Rice University, Houston, TX 77005, USA (e-mail: \mbox{studer@rice.edu}).}
\thanks{The work of C.~Studer was  supported by the Swiss National Science Foundation (SNSF) under Grant PA00P2-134155.}


}


\maketitle


\begin{abstract}

We consider the recovery of sparse signals subject to sparse interference, as introduced in Studer \emph{et al.}, IEEE Trans.\ IT, 2012. 
We present novel probabilistic recovery guarantees for this framework, covering varying degrees of knowledge of the signal and interference support, which are relevant for a large number of practical applications.
Our results assume that the sparsifying dictionaries are 
characterized by coherence parameters and we require randomness only in the signal and/or interference.
The obtained recovery guarantees show that one can recover sparsely corrupted signals with overwhelming probability, even if the sparsity of both the signal and interference scale (near) linearly with the number of measurements.
%
\end{abstract}





\section{Introduction}
\label{sec:intro}
We consider the problem of recovering the sparse signal vector~\mbox{$\sig\in\complexset^{\dima}$} with support set \sigsup (containing the locations of the non-zero entries of \sig) from $\dimm$ linear measurements~\cite{Studer2012}
\begin{align}  \label{eq:signal_model}
    \obs = \dicta \sig + \dictb \err.
\end{align}
Here,  $\dicta\in\complexset^{\dimm\times \dima}$ and $\dictb\in\complexset^{\dimm\times \dimb}$ are given and known  dictionaries, i.e., matrices that are possibly over-complete and whose columns have unit Euclidean norm.
The vector $\err\in\complexset^{\dimb}$ with support set~\errsup represents the sparse interference. 
We investigate the following models for the sparse signal vector~$\sig$ and sparse interference vector $\err$, and their support sets \sigsup and~\errsup:
\begin{itemize}
\item The interference support set  \errsup is \emph{arbitrary}, i.e., $\errsup\subseteq\{1,\ldots,\dimb\}$ can be any subset of cardinality \nerr.
In particular,~\errsup may depend upon the sparse signal vector~$\sig$ and/or the dictionary~$\dicta$, and hence, may also be chosen adversarially. The support set~$\sigsup$ of \sig is chosen  \emph{\rev{uniformly at }random}, i.e., $\sigsup$ is chosen uniformly at random from all subsets of $\{1,\ldots,\dima\}$ with cardinality \nsig.
\item The support set \errsup of the sparse interference vector \err is chosen \rev{uniformly} at random, i.e., $\errsup$ is chosen uniformly at random from all subsets of $\{1,\ldots,\dimb\}$ with cardinality \nerr. The support set~\sigsup is assumed to be arbitrary and of size \nsig.  
\item Both \sigsup and \errsup, the support sets of the signal  and of the interference with size \nsig and \nerr, respectively,  are chosen uniformly at random. 
\end{itemize}
In addition, for each model on the support sets \sigsup and \errsup we may or may not know either of the support sets prior to recovery.
%

%
%


As discussed in~\cite{Studer2012}, recovery of the sparse signal vector \sig from the sparsely corrupted observation $\obs$ in \eqref{eq:signal_model} is relevant in a large number of practical applications. In particular, restoration of saturated \rev{or clipped} signals \cite{Laska2009,Adler2011,Adler2012}, signals impaired by impulse noise \cite{Vaseghi1992,Godsill1998,Novak2010}, or removal of narrowband interference is captured by the input-output relation~\fref{eq:signal_model}. 
Furthermore, the \rev{model}~\fref{eq:signal_model} enables us to investigate sparsity-based super-resolution and in-painting \cite{Elad2001a,Mallat2010}, as well as signal separation~\cite{Elad2005,Cai2009b}.
Hence, identifying the fundamental limits on the recovery of the vector $\sig$ from the sparsely corrupted observation $\obs$ is of significant practical interest.

Recovery guarantees for sparsely corrupted signals have been partially studied in\rev{\cite{Studer2012,Studer2012a,Kuppinger2012,Wright2010,Li2011,McCoy2012,Laska2009,Laska2010,Vaswani2010a,Jacques2010}}. 
%
In particular,~\cite{Studer2012,Studer2012a} investigated coherence-based recovery guarantees for arbitrary support sets \sigsup and \errsup and for varying levels of support-set knowledge; \cite{Kuppinger2012} analyzed the special case where both support sets are unknown, but one is chosen arbitrarily and the other at random.
%
%
The recovery guarantees in\rev{\cite{Wright2010,Li2011,McCoy2012}} require that the measurement matrix \dicta is chosen at random \rev{and that $\dictb$ is unitary.}
The guarantees in  \cite{Laska2009,Laska2010,Vaswani2010a,Jacques2010} characterize \dicta by the restricted isometry property~(RIP), which is, in general, difficult to verify in practice.
%
The recovery guarantees \cite{Li2011,Laska2009,Laska2010} require \dictb to be unitary, whereas  \cite{Vaswani2010a,Jacques2010} only consider a \emph{single} dictionary \dicta and partial support-set knowledge within \dicta.
\rev{The case of support-set knowledge was also addressed in \cite{Vaswani2010b}, but for a model that differs considerably from the setting here. Specifically, \cite{Vaswani2010b} uses a time-evolution model that incorporates support-set knowledge obtained in previous iterations and the corresponding results are based on the RIP.
Finally, \cite{Qiu2011} considered a model where the interference is sparse in an unknown basis.
}
%
%
The specific models and assumptions underlying the results in\rev{\cite{Wright2010,Li2011,McCoy2012,Laska2009,Laska2010,Vaswani2010a,Jacques2010,Vaswani2010b,Qiu2011}} reduce their utility for the applications outlined above.

\subsection{Generality of the signal and interference model}
In this paper, we will exclusively focus on probabilistic results where the randomness is in the signal and/or the interference but \emph{not} in the dictionary. Furthermore, the dictionaries \dicta and~\dictb  will be characterized only by their coherence parameters and their dimensions.
Such results enable us to operate with a given (and arbitrary) pair of sparsifying dictionaries \dicta and~\dictb, rather than hoping that the signal will be sparse in a randomly generated dictionary (as in \cite{McCoy2012}) or that \dicta satisfies the RIP.
The following two application examples illustrate the generality of our results.
\subsubsection{Restoration of saturated signals}
In this example, a signal $\vecy=\dicta\sig$ is subject to saturation~\cite{Studer2012}. 
This impairment is captured by setting $\obs=g_a(\vecy)$ in \fref{eq:signal_model}, where $g_a(\cdot)$ implements element-wise saturation to $[-a,a]$ with $a$ being the saturation level.
By writing $\obs=\vecy + \vece$ with $\vece=g_a(\vecy)-\vecy$, where $\vece$ is non-zero only for the entries where the saturation in~$\obs$ occurs, we see that for moderate saturation levels~$a$, the vector~$\err$ will be sparse. 
The reconstruction of the (uncorrupted) signal~\vecy from the saturated measurement \obs, amounts to recovering \sig from $\obs=\dicta\sig+\err$, followed by computing $\vecy=\dicta\sig$.

We assume that the signal $\vecy=\dicta\sig$ is drawn from a stochastic model where \rev{\sig} has a support set chosen uniformly at random. Since the saturation artifacts modeled by \err are dependent on \vecy, we want to guarantee recovery for  arbitrary \errsup.
Furthermore, we can identify the locations where the saturation occurs (e.g., by comparing the entries of \vecz to the saturation level $a$) and hence, we can assume that~\errsup is known prior to recovery.
The recovery guarantees developed in this paper include this particular combination of support-set knowledge and randomness as a special case, whereas the recovery guarantees in \cite{Studer2012,Kuppinger2012,Tropp2008} are unable to consider all aspects of this model and turn out to be more restrictive.

\subsubsection{Removal of impulse noise}
\label{sec:rin}
Consider a signal $\vecy=\dicta\sig$ that is subject to impulse noise. Specifically, we observe \mbox{$\obs=\vecy+\err$}, where \err is the impulse noise vector. For a sufficiently low impulse-noise rate, \err will be sparse in the identity basis, i.e., $\dictb=\identity$.
As before, consider the setting where $\vecy=\dicta\sig$ is generated from a stochastic model with unknown support set \sigsup.
Since impulse noise does not, in general, depend on the signal~\vecy, we may chose \errsup at random.
In addition, the locations  \errsup of the impulse noise are normally unknown.
%
%

Recovery guarantees for this setting are partially covered by \cite{Studer2012,Kuppinger2012,Tropp2008}. 
However, as for the saturation example above, the recovery guarantees in  \cite{Studer2012,Kuppinger2012,Tropp2008} are unable to exploit all aspects of support-set knowledge and randomness. 
The results developed here cover this particular setting as a special case and hence, lead to less restrictive recovery guarantees.

\rev{In fact, there is an even more general setting compared to \eqref{eq:signal_model}, which encompasses the cases listed in \fref{tab:cases}.  Specifically, a generalization would be to consider the model $\obs=\dicta\sig+\dictb\err$ with $\sigsup=\supp(\sig) = \sigsup_r \cup \sigsup_a$ and $\errsup=\supp(\err) = \errsup_r \cup \errsup_a$ where the support set \sigsup is known and \errsup is unknown, and, furthermore, $\sigsup_a$ and $\errsup_a$ are chosen arbitrarily and $\sigsup_r$ and $\errsup_r$ are chosen uniformly at random.
The analysis of this model, however, is left for future work.\footnote{Note that our model corresponds to the case where two of the sets $\sigsup_r$, $\sigsup_a$, $\errsup_r$, and $\errsup_a$ are forced to be the empty set.}
}

\subsection{Contributions}

In this paper, we present probabilistic recovery guarantees that improve \rev{or refine} 
the ones in\cite{Kuppinger2012,Tropp2008,Studer2012} and cover novel cases for varying degrees of knowledge of the signal and interference support sets. 
Our results depend on the coherence parameters of the two dictionaries \dicta and~\dictb, their dimensions, \rev{and their spectral norms}.
In particular, we present novel recovery guarantees for the situations where the support sets \sigsup and/or \errsup are chosen at random, and for the cases where knowledge of neither, one, or both support sets \sigsup and \errsup is available prior to recovery. 
For the case where one support set is random and the other arbitrary, but no knowledge of $\sigsup$ and $\errsup$ is available, we present an improved (i.e., less restrictive) recovery guarantee than the existing one in \cite[Thm.~6]{Kuppinger2012}.
Finally, we show that $\ell_1$-norm minimization is able to recover the vectors \sig and \err with overwhelming probability, even if the number of non-zero components in both  scales (near) linearly with the number of measurements.

A summary of all the cases studied in this paper is given in \fref{tab:cases}; the theorems highlighted in dark gray indicate novel recovery guarantees, light gray indicates \rev{refined} ones. 
%
%
We will only prove the boldface theorems; the corresponding symmetric cases are shown in italics and the associated recovery guarantees can be obtained by interchanging the roles of  \sig and \err.

\begin{table*}
\centering
\caption{Summary of all recovery guarantees for sparsely corrupted signals.}
\label{tab:cases}
\begin{tabular}{lcccc}
\toprule[0.15em]
& \sigsup, \errsup arbitrary & \sigsup random, \errsup arbitrary & \sigsup arbitrary, \errsup random & \sigsup, \errsup random \\ \midrule[0.1em]
\multirow{2}{*}{\sigsup, \errsup known} & Case \XkaEka & \newresb{Case \XkrEka} & \newresb{\textit{Case \XkaEkr}} & \newresb{Case \XkrEkr}\\ 
 & \oldres{ \cite[Thm.~3]{Studer2012}}  & \newres{\fref{thm:XkrEka}}  &  \newresb{\fref{thm:XkrEka}} & \newres{\fref{thm:XkrEkr}}  \\ \midrule
\multirow{2}{*}{\errsup known} & Case \XuaEka & \newresb{Case \XurEka}  & \newresb{\textit{Case \XuaEkr}} & \newresb{Case \XurEkr} \\ 
& \oldres{\cite[Thm.~4]{Studer2012}} & \newres{Theorem~\ref{thm:XurEka_P0}} &  \newresb{Theorem~\ref{thm:XkrEua_P0} } &  \newres{Theorem~\ref{thm:XurEkr_P0} }\\ \midrule
\multirow{2}{*}{\sigsup known} & \textit{Case \XkaEua} & \newresb{Case \XkrEua} & \newresb{\textit{Case \XkaEur}} & \newresb{\textit{Case \XkrEur}} \\ 
& \oldres{\cite[Cor.~6]{Studer2012}} & \newres{Theorem~\ref{thm:XkrEua_P0}} & \newresb{Theorem~\ref{thm:XurEka_P0} }  &   \newresb{Theorem~\ref{thm:XurEkr_P0}}\\ \midrule
\multirow{2}{*}{neither known} & Case \XuaEua & \improvresb{Case \XurEua} & \improvresb{\textit{Case \XuaEur}} & \newresb{Case \XurEur}\\ 
& \oldres{\cite[Thms.~2 and 3]{Kuppinger2012}} & \improvres{Theorem~\ref{thm:XurEua_P0} and \cite[Thm.~6]{Kuppinger2012} } & \improvresb{Theorem~\ref{thm:XurEua_P0} and \cite[Thm.~6]{Kuppinger2012} } & \newres{Theorem~\ref{thm:XurEur_P0} }\\
\bottomrule[0.15em]
\end{tabular}
\end{table*}

\subsection{Notation}
\label{sec:notation}
Lowercase and uppercase boldface letters stand for column vectors and matrices, respectively. For the matrix~\bM, we denote its transpose, adjoint, and  (Moore--Penrose) pseudo-inverse by $\bM^T$, $\herm{\bM}$, and $\bM^{\dagger}$, respectively. The $j$th column and the entry in the $i$th row and $j$th column of the matrix~$\bM$ is designated by $\bmm_j$ and $[\bM]_{i,j}$, respectively. 
%
%
%
%
%
The minimum and maximum singular value of $\bM$ are given by $\smin(\bM)$ and $\smax(\bM)$, respectively; the spectral norm is $\norm{l2l2}{\bM}={\smax(\bM)}$. 
The $\ell_1$-norm of the vector $\bmv$ is denoted by $\norm{l1}{\bmv}$ and $\norm{l0}{\bmv}$ stands for the number of nonzero entries in \vecv. 
%
%
Sets are designated by upper-case calligraphic letters; the cardinality of the set \setS is $\abs{\setS}$. 
The support set of \vecv, i.e., the indices of the  nonzero entries, is given by $\supp(\vecv)$. 
%
%
%
%
%
%
%
The matrix $\bM_\setS$ is obtained from \bM by retaining the columns of \bM with indices in~$\setS$; the vector $\bmv_\setS$ is obtained analogously from the vector \bmv. 
The $\sign(\cdot)$ function applied to a vector returns a vector consisting of the phases of each entry.
The $N\times N$ restriction matrix~$\bR_\setS$ for the set $\setS\subseteq\{1,\ldots,N\}$ has $[\bR_\setS]_{k,k} = 1$ if  $k\in\setS$ and is zero otherwise.
%
%
For random variables~$X$ and~$Y$, we define $\moment{q}{X} \define \expected{\abs{X}^q}^{1/q}$ to be the $q$th moment\rev{, which defines an $\ell_q$-norm on the space of complex-valued random variables, and hence satisfies the triangle inequality.}  \rev{We define} $\momentx{q}{X}{f(X,Y)}$ to be the $q$th moment with respect to $X$ and we define $\isonb$ to be equal to $1$ if the condition $\mu\neq 0$ holds and~$0$ otherwise.
\rev{For two functions $f$ and $g$ we write $f\sim g$ to indicate that $f(n)/g(n)\rightarrow 1$ as $n\rightarrow \infty$, and we say that ``$f$ scales with~$g$.''}

Throughout the paper, $\sigsup=\supp(\sig)$ is assumed to be of cardinality \nsig and $\errsup=\supp(\err)$ of cardinality~\nerr.
We define $\dict = [\, \dicta\, \, \dictb \,]$ and  $\dictse = [\, \dicta_\sigsup\,\, \dictb_\errsup\,]$ to be the sub-dictionary of~$\dict$ associated with the non-zero entries of \sig and \err.
%
Similarly, we define the vector $\ssigse = [\,\sig^\trans_\sigsup\,\,\err^\trans_\errsup\,]^T$  which consists of the non-zero components of $\ssig = [\,\sig^\trans\,\err^\trans\,]^T$.
%

\subsection{Outline of the paper}

The remainder of the paper is organized as follows. Related prior work is summarized in \fref{sec:background}. The main theorems are presented in \fref{sec:results} and a corresponding discussion is given in \fref{sec:discussion}. We conclude in \fref{sec:conc}. All proofs are relegated to the Appendices.


\section{Related Prior Work}
\label{sec:background}

We next summarize relevant prior work on sparse signal recovery and sparsely corrupted signals, and we put our results into perspective.

\subsection{Coherence-based recovery guarantees}
During the last decade, numerous deterministic and probabilistic guarantees for the recovery of sparse signals from linear (and non-adaptive) measurements have been developed \cite{Donoho2003,Tropp2004,Chen1998,Candes2007,Tropp2008,Candes2010d,Candes2006b,Candes2008d,Cai2010a}.
These results give sufficient conditions for when one can reconstruct the sparse signal vector \sig from the (interference-less) observation $\vecy=\dicta\sig$ by solving
\begin{align*}
    \PZ \quad \underset{\hat\sig}{\text{minimize}} \norm{l0}{\hat{\sig}} \quad \text{subject to} \,\, \vecy = \dicta\hat{\sig},
\end{align*}
or its convex relaxation, known as basis pursuit, defined as
\begin{align*}
   \BP \quad \underset{\hat{\sig}}{\text{minimize}} \norm{l1}{\hat{\sig}} \quad \text{subject to} \,\, \vecy = \dicta\hat{\sig}.
\end{align*}
%
In particular, in~\cite{Donoho2003,Chen1998,Tropp2004}  it is shown that if $\norm{l0}{\sig}\le\nsig$ for some $\nsig < \left(1+1/{\coha}\right)/2$ with the coherence parameter 
\begin{align} \label{eq:coha}
    \coha \define \max_{i,j,i\neq j} \, \abs{\scalprod{}{\cola_i}{ \cola_j}},
\end{align}
then \PZ and \BP are able to perfectly recover the sparse signal vector~$\sig$.
%
Such coherence-based recovery guarantees are, however, subject to the 
``square-root bottleneck'', which only guarantees the recovery of  $\vecx$ for sparsity levels on the order of $\nsig \sim  \sqrt{\dimm}$~\cite{Tropp2008}.
This behavior is an immediate consequence of the Welch bound~\cite{Welch1974} and dictates that the number of measurements must grow at least quadratically in the sparsity level of $\vecx$ to guarantee  recovery. 
In order to overcome this square-root bottleneck, one must either resort to a RIP-based analysis, e.g.,~\cite{Candes2007,Candes2010d,Candes2006b,Candes2008d}, which typically requires randomness in the dictionary \dicta, or a \emph{probabilistic} analysis that only considers randomness in the vector \vecx, \rev{and where} $\dicta$ is constant (known) and solely characterized by its coherence parameter\rev{, dimension, and spectral norm}~\cite{Tropp2008}.
In this paper, we are interested in the latter type of results. 
%
%
Such probabilistic and coherence-based recovery guarantees that overcome the square-root bottleneck have been derived for \PZ and \BP in~\cite{Tropp2008}. The corresponding results, however, do not exploit the structure of the problem \fref{eq:signal_model}, i.e., the fact that we are dealing with two dictionaries and that knowledge of~\sigsup and/or \errsup may be available prior to recovery.

\subsection{Recovery guarantees for sparsely corrupted signals}

Guarantees for the recovery of sparsely corrupted signals as modeled by \eqref{eq:signal_model} have been developed recently in \cite{Studer2012,Studer2012a,Kuppinger2012}.
%
%
The reference \cite{Studer2012} considers deterministic (and coherence-based) results for several cases\footnote{Note that no efficient recovery algorithm with corresponding guarantees is known for the case studied in~\cite{Studer2012}, where only the cardinality of \sigsup or \errsup is known. Thus, we do not consider this case in the remainder of the paper.} which arise in different applications:
\begin{inparaenum}[1)]
\item $\sigsup=\supp(\sig)$ and $\errsup = \supp(\err)$ are known prior to recovery,
\item only one of \sigsup and \errsup is known, and
\item neither \sigsup nor \errsup are known. 
\end{inparaenum}
For case 1), the non-zero entries of both the signal and interference vectors can be recovered by \cite{Studer2012}
\begin{align}
\ssig_{\sigsup,\errsup}=\pinv{\dictse} \obs, \label{eq:both_conc}
\end{align}
if the recovery guarantee in \cite[Thm.~2]{Studer2012} is satisfied.
For case 2), recovery is performed by using modified versions of \PZ and \BP; the associated recovery guarantees can be found in \cite[Thm.~4 and Cor.~6]{Studer2012}.
For case 3), recovery guarantees for the standard \PZ or \BP algorithms are given in \cite[Thms.~2 and 3]{Kuppinger2012}.
However, all these recovery guarantees suffer from the square-root bottleneck, as they guarantee recovery for \emph{all} signal and \emph{all} interference vectors satisfying the given sparsity constraints. 
A notable exception for case 3) was discussed in \cite[Thm.~6]{Kuppinger2012}. There, \err is assumed to be random, but $\sig$ is assumed to be arbitrary. This model overcomes the square-root bottleneck and is able to significantly improve upon the corresponding deterministic recovery guarantees in \cite[Thms.~2 and 3]{Kuppinger2012}.

Another strain of recovery guarantees for sparsely corrupted signals that are able to overcome the square-root bottleneck have been developed in~\cite{Li2011,McCoy2012,Laska2009,Laska2010,Vaswani2010a,Jacques2010,Wright2010}. 
The \rev{references~\cite{Wright2010,Li2011,McCoy2012}} consider the case where $\dicta$ is random, whereas~\cite{Laska2009,Laska2010,Vaswani2010a,Jacques2010} consider matrices $\dicta$ that are characterized by the RIP, which is, in general, difficult to verify for a given (deterministic)~\dicta.
\rev{Indeed, it has been recently shown that calculating the RIP for a given matrix is NP-hard \cite{Pfetsch2012}.}
Moreover, the recovery guarantees in\rev{\cite{Wright2010,Li2011,Laska2009,McCoy2012,Laska2010}} require that $\dictb$ is an orthogonal matrix and, hence, these results do not allow for \emph{arbitrary} pairs of dictionaries $\dicta$ and $\dictb$. 
In addition, \cite{Li2011,Laska2010} do not study the impact of support-set knowledge on the recovery guarantees.
The results in \cite{Vaswani2010a,Jacques2010} only consider a \emph{single} dictionary with partial support-set knowledge and, thus, are unable to exploit the fact that the signal and interference exhibit sparse representations in two \emph{different} dictionaries.
%
%
%
While all these assumptions are valid for applications based on compressive sensing (see, e.g.,  \cite{Candes2005a,Donoho2006}), they are not suitable for the application scenarios outlined in \fref{sec:intro}.

To overcome the square-root bottleneck for \emph{arbitrary} pairs of dictionaries $\dicta$ and $\dictb$, we next \rev{propose} a generalization of the probabilistic models developed in  \cite{Kuppinger2012,Tropp2008} for the cases 1), 2), and 3) outlined above.
In particular, we impose a random model on the signal and/or interference vectors rather than on the dictionaries, and we allow for varying degrees of knowledge of the support sets \sigsup and \errsup.
An overview of the coherence-based recovery guarantees developed next is given in \fref{tab:cases}.


\section{Main Results}
\label{sec:results}

\begin{model}[t]
\caption{\pzeromodel}
\label{mod:p0}
\begin{itemize}
\item Let $\sig\in\complexset^\dima$ and $\err\in\complexset^\dimb$ have support set \sigsup and \errsup, respectively, of which at least one is chosen \rev{uniformly} at random \rev{and where the non-zero entries of both \sig and \err are drawn from a continuous distribution.} 
\item The observation \obs is given by $\obs = \dicta\sig + \dictb \err $.
\end{itemize}
\end{model}

\begin{model}[t]
\caption{\ponemodel}
\label{mod:p1}
\begin{itemize}
\item  The conditions of \pzeromodel hold.
\item If \sigsup or \errsup is \rev{unknown}, then assume that the corresponding non-zero entries of the associated vector(s) are drawn from a continuous distribution, where the phases of the individual components are independent and uniformly distributed on $[0,2\pi)$.
\end{itemize}
\end{model}

The recovery guarantees developed next rely upon the models \pzeromodel and \ponemodel summarized in Model~1 and Model~2, respectively. \rev{Model~2 differs subtly from the model in \cite{Kuppinger2012} in that we do not require the uniform phase assumption in the vector with known support, a setting which was not considered in \cite{Kuppinger2012}.}
\rev{In addition to the Models~1 and~2, our results require} 
 the coherence parameters\footnote{\rev{Note that we could also characterize the dictionaries \dicta and \dictb with the cumulative coherence \cite{Tropp2004}.  For the sake of simplicity of exposition, however, we stick to the  coherence parameters \coha, \cohb, and \mcoh only.}} of the dictionaries~$\dicta$ and \dictb, i.e., the coherence \coha of \dicta in~\eqref{eq:coha}, the coherence \cohb of~\dictb given by
\begin{align*} 
\cohb \define \max_{i,j,i\neq j} \, \abs{\scalprod{}{\colb_i}{ \colb_j}},
\end{align*}
and the mutual coherence \mcoh between \dicta and \dictb, defined as
\begin{align*}
  \mcoh = \max_{i,j} \, \abs{\scalprod{}{\cola_i} {\colb_j}}.
\end{align*}
Our main results for the cases highlighted in \fref{tab:cases} are detailed next.

\subsection{Cases  \XkrEka and \XkrEkr:~\sigsup and \errsup  known}
\label{sec:both_known}

We start with the case where both support sets~\sigsup and~\errsup are known prior to recovery. 
The following theorem guarantees recovery of $\vecx$ and $\vece$ from $\vecz$, \rev{using~\fref{eq:both_conc}}, with high probability.


\begin{thm}[Cases \XkrEka and \XkrEkr] \label{thm:XkrEka}  \label{thm:XkrEkr}
Let $\sig$ and $\err$ be signals satisfying the conditions of \pzeromodel, assume that both \sigsup and \errsup are known, and choose $\beta \ge\log(\nsig)$.
If \sigsup is chosen uniformly at random, \errsup is arbitrary, and if 
\begin{align}
\delta e^{{-\frac 1 4}} \ge & \, \norm{l2l2}{\dicta} \norm{l2l2}{\dictb} \sqrt{\frac{\nsig}{\dima}} + 12 \coha \sqrt{\beta \nsig} +  (\nerr-1)\cohb \nonumber\\
&+\isonba\frac{2\nsig}{\dima} \norm{l2l2}{\dicta}^2 +3\mcoh \sqrt{2\beta \nerr}, \label{eq:smin_ra}
\end{align}
holds with\footnote{\rev{Later we will require \eqref{eq:smin_ra} to hold for different values of $\delta$.}} $\delta = 1$, then we can recover $\sig$ and $\err$ using~\fref{eq:both_conc} with probability at least $1-e^{-\beta}$.

If both \sigsup and \errsup are chosen at random and if 
\begin{align}
\delta e^{{-\frac 1 4}} \ge & \ 12 \sqrt{\beta}\left(\coha \sqrt{ \nsig} +  \cohb \sqrt{ \nerr}\right) +\isonba\frac{2\nsig}{\dima} \norm{l2l2}{\dicta}^2   +\isonbb\frac{2\nerr}{\dimb} \norm{l2l2}{\dictb}^2  \nonumber\\
&+ \min \!\bigg\{ 3\mcoh \sqrt{2\beta\nsig}   +   \sqrt{\frac{\nerr}{\dimb}} \norm{l2l2}{\dicta^\adj \dictb}, 3\mcoh \sqrt{2\beta\nerr} + \sqrt{\frac{\nsig}{\dima}} \norm{l2l2}{\dicta^\adj \dictb}\!\bigg\},
\label{eq:smin_rr}
\end{align}
holds with $\delta=1$ and $\beta \ge \max\{\log(\nsig),\log(\nerr)\}$, then we can recover $\sig$ and $\err$ using~\fref{eq:both_conc} with probability at least $1-e^{-\beta}$.
\end{thm}
\begin{IEEEproof}
See \fref{app:XEknown}. 
\end{IEEEproof}

A discussion of the recovery conditions \fref{eq:smin_ra} and \fref{eq:smin_rr} is relegated to \fref{sec:discussion}.\rev{\footnote{\rev{In order to slightly improve the conditions in \eqref{eq:smin_ra} or \eqref{eq:smin_rr}, one could replace the term $(\nerr-1)\cohb$ with the cumulative coherence as defined in~\cite{Tropp2004}. 
}}}

\subsection{Cases \XurEka and \XurEkr:~\setE known}
\label{sec:Eknown}
Consider the case where only the support set  $\errsup$ of $\err$ is known prior to recovery.  
In this case, recovery of \vecx (and the non-zero entries of \err) from \vecz can be achieved by solving~\cite{Studer2012}\footnote{\rev{Note that since $\errsup$ is known, the term $\norm{l0}{\hat{\err}_\errsup}$  in $\PzeroEs$ can be omitted.  We keep the term, however, for the sake of consistency with the problem \PoneEs.}}
\begin{align}
\PzeroEs \quad \left\{\begin{array}{ll} \underset{\hat\sig, \hat\err_\errsup}{\mini} & \norm{l0}{\hat{\sig}} \rev{+\norm{l0}{\hat{\err}_\errsup} } \\ 
       \st & \obs = \dicta \hat{\sig}  + \dictb_\errsup\hat\err_\errsup,
\end{array}\right.
\end{align}
or its convex relaxation\footnote{\rev{Note that we consider a slightly different convex optimization problem \PoneEs to that proposed in \cite{Studer2012}, \PoneE, for the case where $\errsup$ is known prior to recovery.  In practice, however, both problems exhibit similar recovery performance.}}
\begin{align}
    \PoneEs \quad \left\{\begin{array}{ll} \underset{\hat\sig, \hat\err_\errsup}{\mini} & \norm{l1}{\hat{\sig}} \rev{+ \norm{l1}{\hat{\err}_\errsup}}\\ 
       \st & \obs = \dicta \hat{\sig}  + \dictb_\errsup\hat\err_\errsup.
\end{array}\right.
\end{align}
The following theorems guarantee the recovery of $\vecx$ and \err from \vecz, using \PzeroEs or \PoneEs, with high probability.


\begin{thm}[Case \XurEka] \label{thm:XurEka_P0}  \label{thm:XurEka_BP}
Let $\sig$ and~$\err$ be signals satisfying the conditions of \pzeromodel, assume that~\errsup is known prior to recovery and chosen arbitrarily, and  assume that \sigsup is unknown and drawn uniformly at random.  
Choose $\beta\ge\log(\nsig)$. 
If \eqref{eq:smin_ra} holds for some $\delta$ where $0<\delta<1$ and if
\begin{align}
\nsig \coha^2 + \nerr \mcoh^2 & < 1-\delta,\label{eq:nxa_nem_P0}
\end{align}
then we can recover \sig and \err using $\PzeroEs$ with probability at least  $1-e^{-\beta}$. 

Moreover, if $\sig$ and $\err$ are  signals satisfying the conditions of \ponemodel,  and, in addition to \fref{eq:smin_ra}, 
\rev{if}
\begin{align}
     \nsig \coha^2 + \nerr \mcoh^2 & < \frac{(1-\delta)^2}{2(\log(\dima) + \beta)}, \label{eq:nxa_nem_BP}
 \end{align}
holds, 
then we can recover \sig and \err using  $\PoneEs$ with probability at least $1-3e^{-\beta}$.
\end{thm}

\begin{IEEEproof}
See Appendices \ref{app:PZ} and \ref{app:bp}.
\end{IEEEproof}

Note that 
%
by combining \eqref{eq:smin_ra}, \eqref{eq:nxa_nem_P0}, and possibly \eqref{eq:nxa_nem_BP} into a \emph{single} recovery condition, thereby effectively removing~$\delta$, we can easily calculate the largest values of \nsig and \nerr for which successful recovery with high probability is guaranteed (see \fref{sec:asymptconditions} for a corresponding discussion).

\begin{thm}[Case \XurEkr] \label{thm:XurEkr_P0}  \label{thm:XurEkr_BP}
Let $\sig$ and $\err$ be signals satisfying the conditions of \pzeromodel, assume that \errsup is known but \sigsup is unknown prior to recovery, and  assume that both \sigsup  and \errsup are drawn uniformly at random.
If  \eqref{eq:smin_rr} and \eqref{eq:nxa_nem_P0} hold 
for some $0<\delta<1$ and $\beta \ge \max\{\log(\nsig),\log(\nerr)\}$,
then we can recover $\sig$ and $\err$ using \PzeroEs with probability at least $1-e^{-\beta}$.

Moreover, if \sig and \err are signals satisfying the conditions of \ponemodel and if \eqref{eq:nxa_nem_BP}
holds in addition to \eqref{eq:smin_rr} and \eqref{eq:nxa_nem_P0}, then we can recover~$\sig$ and~$\err$ using \PoneEs with probability  at least $1-3e^{-\beta}$.
\end{thm}
\begin{IEEEproof}
See Appendices \ref{app:PZ} and \ref{app:bp}.
\end{IEEEproof}

A discussion of both theorems is relegated to \fref{sec:discussion}.


\subsection{Case \XkrEua:~\sigsup known}
The case where \sigsup is random and known, and \errsup is unknown and arbitrary, differs slightly to the case where \sigsup is random and unknown, and \errsup is arbitrary and known (covered by \fref{thm:XurEka_P0}). Hence, we need to consider both cases separately.
The recovery problems \PzeroXs and \PoneXs required here are defined analogously to \PzeroEs and \PoneEs.

\begin{thm}[Case \XkrEua] \label{thm:XkrEua_P0}
Let $\sig$ and $\err$ be signals satisfying the conditions of \pzeromodel, assume that the support set \sigsup is known and chosen uniformly at random, and assume that \errsup is unknown and arbitrary.
If  
\begin{align}
\delta e^{{-\frac 1 4}} \ge & \norm{l2l2}{\dicta} \norm{l2l2}{\dictb} \sqrt{\frac{\nerr}{\dimb}} + 12 \cohb \sqrt{\beta \nerr}  +  (\nsig-1)\coha \nonumber\\
&+\isonbb\frac{2\nerr}{\dimb} \norm{l2l2}{\dictb}^2+3\mcoh \sqrt{2\beta \nsig}, \label{eq:smin_ar}
\end{align}
holds for some $0<\delta<1$ and $\beta \ge \log(\nerr)$, and if 
\begin{align}
     \nsig \mcoh^2 + \nerr \cohb^2 & < 1-\delta, \label{eq:nxm_neb_P0}
 \end{align}
then  we can recover \sig and \err using $\PzeroX$ with probability at least $1-e^{-\beta}$.

Moreover, if $\sig$ and $\err$ are signals satisfying the conditions of \ponemodel, and, in addition to \fref{eq:smin_ar}, \rev{if} 
 \begin{align}
    \nsig \mcoh^2 + \nerr \cohb^2 & < \frac{(1-\delta)^2}{2(\log(\dimb) + \beta)}, \label{eq:nxm_neb_BP} 
\end{align}
holds, 
then we can recover \sig and \err using $\PoneX$ with probability at least $1-3e^{-\beta}$.
\end{thm}
\begin{IEEEproof}
See Appendices \ref{app:PZ} and \ref{app:bp}.
\end{IEEEproof}

A discussion of this theorem is relegated to \fref{sec:discussion}.

\subsection{Cases \XurEua and \XurEur:~No support-set knowledge}
Recovery guarantees for the case of no support-set knowledge, but where one support set is chosen at random and the other arbitrarily can be found in \cite[Thm.~6]{Kuppinger2012}. 
The theorem shown next \rev{is able to refine the result in} \cite[Thm.~6]{Kuppinger2012}.
The  \rev{refinements} are due to the following facts:~i) We allow for arbitrary~$0<\delta<1$, whereas $\delta=1/2$ in \cite[Thm.~6]{Kuppinger2012}, \rev{ii) we add a correction term improving the bounds when either \dicta or \dictb are unitary, }and iii) we do not use a global coherence parameter $\coh = \max\{\coha,\cohb,\mcoh\}$, but rather we further exploit the individual coherence parameters \coha, \cohb, and~\mcoh of \dicta and \dictb. See \fref{sec:recguaranteesexplained} for a corresponding discussion.

\begin{thm}[Case \XurEua] \label{thm:XurEua_P0} \label{thm:XurEua_BP}
Let $\sig$ and $\err$ be signals satisfying the conditions of \pzeromodel,  assume that \sigsup is chosen uniformly at random, and assume that \errsup is arbitrary.
If \eqref{eq:smin_ra}, \eqref{eq:nxa_nem_P0}, and \eqref{eq:nxm_neb_P0} hold 
for some $0<\delta<1$ and $\beta \ge \log(\nsig)$, 
then  
\begin{align*}
    \PZC \quad \underset{\hat\vecx,\hat\vece}{\text{minimize}} \norm{l0}{\hat{\vecx}}+\norm{l0}{\hat{\vece}} \quad \text{subject to} \,\, \vecz = \dicta\hat\vecx+\dictb\hat\vece,
\end{align*}
recovers $\vecx$ and $\vece$ with probability at least $1-e^{-\beta}$.

Moreover, if \sig and \err are  signals satisfying the conditions of \ponemodel and if \eqref{eq:nxa_nem_BP} and \eqref{eq:nxm_neb_BP}
hold in addition to \eqref{eq:smin_ra}, \eqref{eq:nxa_nem_P0}, and \eqref{eq:nxm_neb_P0}, then 
\begin{align*}
   \BPC \quad \underset{\hat\vecx,\hat\vece}{\text{minimize}} \norm{l1}{\hat{\vecx}}+\norm{l1}{\hat{\vece}} \quad \text{subject to} \,\, \vecz = \dicta\hat\vecx+\dictb\hat\vece,
\end{align*}
recovers $\vecx$ and $\vece$  with probability at least $1-3e^{-\beta}$. 
\end{thm}
\begin{IEEEproof}
See Appendices \ref{app:PZ} and \ref{app:bp}
\end{IEEEproof}

\begin{thm}[Case \XurEur]\label{thm:XurEur_P0}\label{thm:XurEur_BP}
Let $\sig$ and $\err$ be signals satisfying the conditions of \pzeromodel and assume that \sigsup and \errsup are both unknown and chosen uniformly at random.
If \eqref{eq:smin_rr}, \eqref{eq:nxa_nem_P0}, and \eqref{eq:nxm_neb_P0} hold 
for some $0<\delta<1$ and $\beta \ge \max\{\log(\nsig),\log(\nerr)\}$, 
then $\PZC$ recovers $\vecx$ and $\vece$ with probability at least $1-e^{-\beta}$.

Moreover, if \sig and \err are signals from \ponemodel and if  \eqref{eq:nxa_nem_BP} and \eqref{eq:nxm_neb_BP}
hold in addition to \eqref{eq:smin_rr}, \eqref{eq:nxa_nem_P0}, and \eqref{eq:nxm_neb_P0}, then~$\BPC$ recovers $\vecx$ and $\vece$ with probability at least $1-3e^{-\beta}$.
\end{thm}
\begin{IEEEproof}
See Appendices \ref{app:PZ} and \ref{app:bp}.
\end{IEEEproof}

A discussion of both theorems is given below.


\section{Discussion of the Recovery Guarantees}
\label{sec:discussion}

We now discuss the theorems presented in \fref{sec:results}. In particular, we study the impact of support-set knowledge on the recovery guarantees and characterize the asymptotic behavior of the corresponding recovery conditions, i.e., the threshold for which recovery is guaranteed with high probability. 

In the ensuing discussion, \rev{we consider two scenarios.  For the first scenario,} we assume that \dicta and \dictb are unitary, i.e., $\dima=\dimb=\dimm$ and \mbox{$\coha=\cohb=0$}, and maximally incoherent, i.e., $\mcoh = 1/\sqrt{\dimm}$. For example, \dicta  could be the discrete Fourier transform (or Hadamard) matrix with appropriately normalized columns and~\dictb the identity matrix.  \rev{The corresponding plots are shown in \fref{fig:unitary}.}
\rev{For the second scenario, \dicta is assumed to be unitary and \dictb is assumed to be the concatenation of two unitary matrices so that $\dimm=\dima=10^8$, $\dimb = 2\dima$, $\coha=0$, and $\cohb=\mcoh = 1/\sqrt{\dimm}$ as described in \cite{Calderbank1997,Gribonval2003a}.  The corresponding plots are shown in \fref{fig:3onb}.}
\rev{In each case we} set $\beta = \log(\dimm)$ or \rev{$\beta = \log(\dimm)/3$ for the $\ell_0$-norm and $\ell_1$-norm-based recovery problems, respectively}, so that recovery is guaranteed with probability at least $1-1/\dimm$.

In order to plot the recovery conditions, we note that for a pair of unitary matrices and a given~\nerr, the recovery conditions of the theorems are quadratic equations in $\sqrt{\nsig}$; this enables us to calculate the maximum \nsig guaranteeing the successful recovery of \vecx and \vece in closed form.

\subsection{Recovery guarantees}
\label{sec:recguaranteesexplained}

\subsubsection{\sigsup and \errsup known}

\begin{figure*}[t]
\centering
  \begin{subfigure}[t]{0.32\textwidth}
     \centering
    \includegraphics[height=1.05\textwidth]{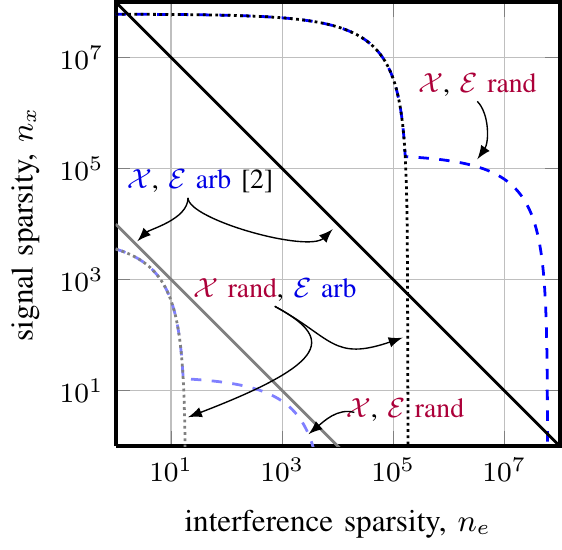}

    \caption{\sigsup and \errsup known  }
    \label{fig:both_known}
  \end{subfigure}
 \begin{subfigure}[t]{0.32\textwidth}
        \centering
 \includegraphics[height=1.05\textwidth]{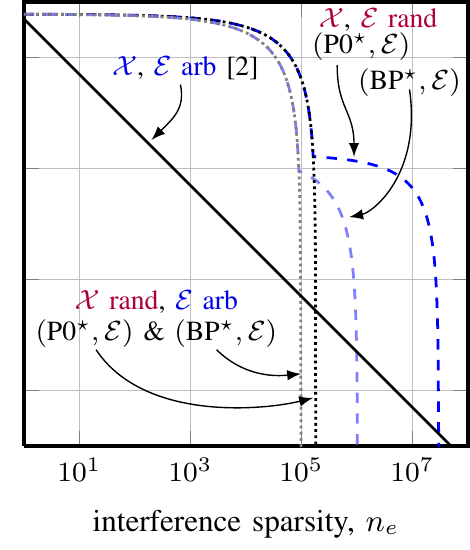}
    \caption{ \errsup  known  }
    \label{fig:one_known}
  \end{subfigure}\hspace{-0.7cm}
 \begin{subfigure}[t]{0.32\textwidth}
   \centering
    \includegraphics[height=1.05\textwidth]{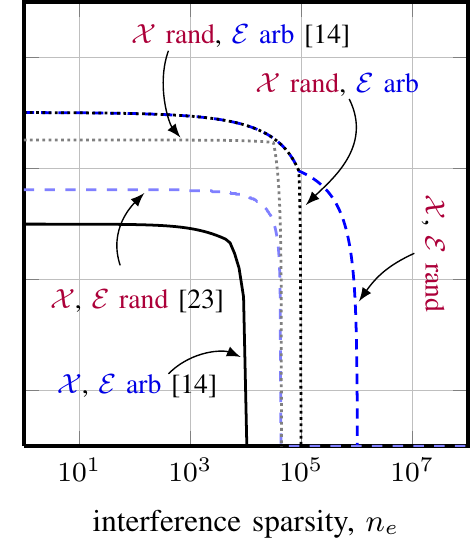}
    \caption{\sigsup and \errsup unknown   }
    \label{fig:none_known}

 \end{subfigure}
 \caption{\dicta and \dictb are \rev{assumed to be} unitary with $\dimm=\dima=\dimb=10^8$ and $\mcoh=1/\sqrt{\dimm}$. In (a) the darker curves in the upper-right are for $m=10^8$ and the  lighter curves in the lower-left are for $m=10^4$. In (c) we show the recovery regions only for \BPC. In each case, recovery is guaranteed with probability at least $1-10^{-8}$. }
\label{fig:unitary}
\end{figure*}

\fref{fig:both_known} shows the recovery conditions  for the cases when both support sets \sigsup and \errsup are assumed to be known.  
For small problem dimensions, i.e., $\dimm=10^4$, the recovery conditions where both support sets are assumed to be arbitrary turn out to be less restrictive than for the case where both support sets are chosen at random.
For large problem dimensions, i.e., $\dimm=10^8$, we see, however, that the probabilistic results of \fref{thm:XkrEkr} guarantee the recovery (with high probability) for larger \nsig and \nerr than the deterministic results of~\cite{Studer2012} considering arbitrary support sets.
Hence, the probabilistic recovery conditions presented here require a sufficiently large problem size in order to  outperform the corresponding deterministic results.
We furthermore see from \fref{fig:both_known} that one can guarantee the recovery of signals having a larger number of non-zero entries if both support sets are chosen at random compared to the situation where~\sigsup is random but~\errsup is arbitrary.

\subsubsection{Only \errsup known}
\fref{fig:one_known} shows the recovery conditions from Theorems \ref{thm:XurEka_BP} and \ref{thm:XurEkr_BP} for the cases where only \errsup is known prior to recovery (the case of only \sigsup known behaves analogously). 
We see that for a random \sigsup and random \errsup successful recovery at high probability is guaranteed for significantly larger \nsig and \nerr compared to the case where one or both support sets are assumed to be arbitrary.
Hence, having more randomness in the support sets leads to less restrictive recovery guarantees. 
%
\rev{We now see that the recovery conditions for \PzeroEs are slightly less restrictive than those for~\PoneEs.}

\subsubsection{No support-set knowledge}
Finally, \fref{fig:none_known} shows the recovery conditions for  \BPC  for the case of no support-set knowledge.
We see that for random  \sigsup and \errsup, successful recovery is guaranteed for significantly larger \nsig and \nerr compared to the case where one or both support sets are assumed to be arbitrary.
As a comparison, we also show the recovery conditions derived in \cite[Thm.~6]{Kuppinger2012} and the conditions from \cite{Tropp2008}, the latter of which does not take into account the structure of the problem \fref{eq:signal_model}.
We see that the recovery conditions derived in Theorems~\ref{thm:XurEua_BP} and~\ref{thm:XurEur_BP}  are less restrictive, i.e., they guarantee the successful recovery (with high probability) for a larger number of nonzero coefficients in both the sparse signal vector~\sig and the sparse interference~\err.

\subsubsection{Non-unitary \dictb}
\begin{figure*}[t]
  \centering
  \begin{subfigure}{0.32\textwidth}
    \centering
        \includegraphics[height=1.05\textwidth]{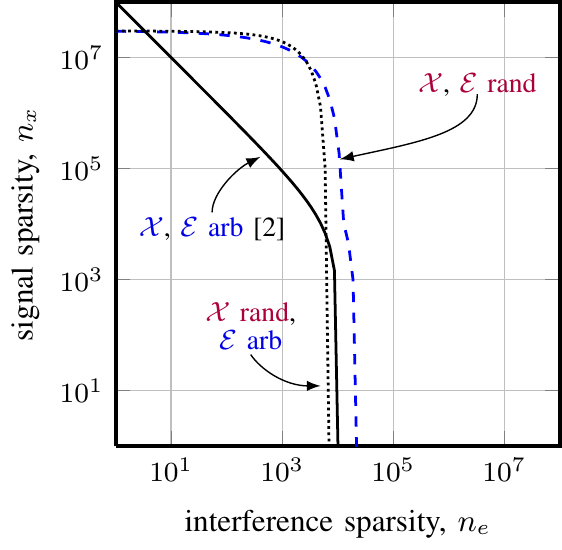}

    \caption{\sigsup and  \errsup  known }
    \label{fig:both_known2}
  \end{subfigure}
  \begin{subfigure}{0.32\textwidth}
    \centering
        \includegraphics[height=1.05\textwidth]{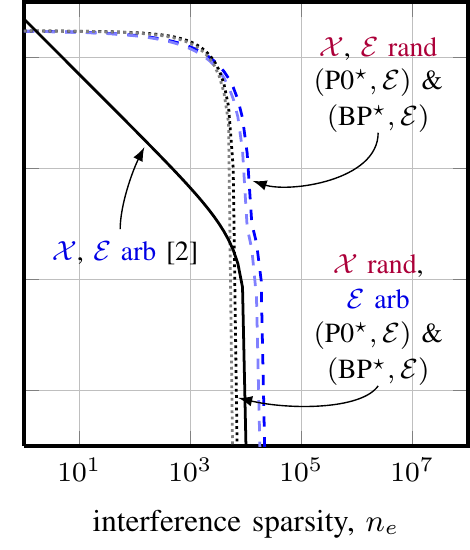}

      \caption{ \errsup  known}
      \label{fig:one_known2}
    \end{subfigure}  \hspace{-0.8cm}
    \begin{subfigure}{0.32\textwidth}
      \centering
             \includegraphics[height=1.05\textwidth]{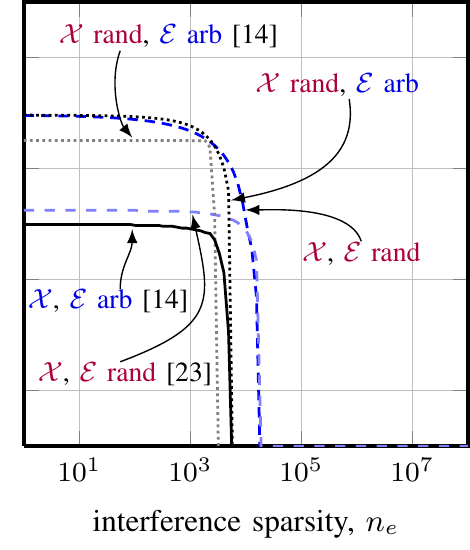}

        \caption{\sigsup and \errsup unknown}
        \label{fig:none_known2}
      \end{subfigure}
      \caption{\dicta is assumed to be unitary and \dictb is assumed to be the concatenation of two unitary matrices so that $\dimm=\dima=10^8$, $\dimb = 2\dima$, $\coha=0$, and $\cohb=\mcoh = 1/\sqrt{\dimm}$ as described in \cite{Calderbank1997,Gribonval2003a}.  In (c) we show the recovery regions only for \BPC. In each case, recovery is guaranteed with probability at least $1-10^{-8}$.}
      \label{fig:3onb}
    \end{figure*}

\rev{
We now consider the setting where \dictb is the concatenation of two unitary matrices and plot the corresponding recovery threshold for differing levels of support set knowledge in \fref{fig:3onb}.
For a fixed \nsig and \dima, we see that by increasing \dimb and \cohb, we suffer a significant loss in the number of non-zero entries of \err that we can recover, when compared to the case where \dictb is unitary.  However, the number of non-zero entries of \sig that we can guarantee to recover is virtually unchanged---an effect which is also present in the deterministic recovery conditions \cite{Studer2012}. 
}

\subsection{Impact of support-set knowledge}

As detailed in \cite{Studer2012}, having knowledge of the support set of \sig or \err implies that one can guarantee the recovery of \sig and \err having up to twice as many non-zero entries (compared to the case of no support-set knowledge).

\begin{figure}[t]
    \centering
    \includegraphics[height=8cm]{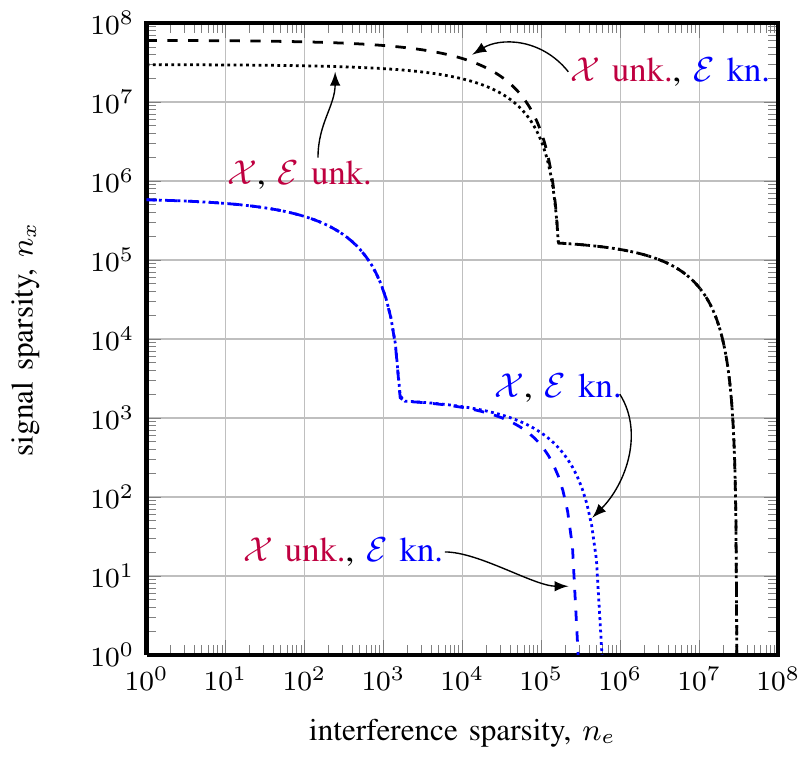}

    \caption{Impact of support-set knowledge on the recovery conditions for \rev{\eqref{eq:both_conc},} \PZ, and \PzeroEs in the case where \sigsup and \errsup are both random. \dicta and \dictb are  unitary with $\dimm=\dima=\dimb=10^6$ (lower-left curves) and $\dimm=\dima=\dimb=10^8$ (upper-right curves) and $\mcoh=1/\sqrt{\dimm}$.}
    \label{fig:knowledge}
\end{figure}

\begin{figure}[t]
    \centering
       \includegraphics[height=8cm]{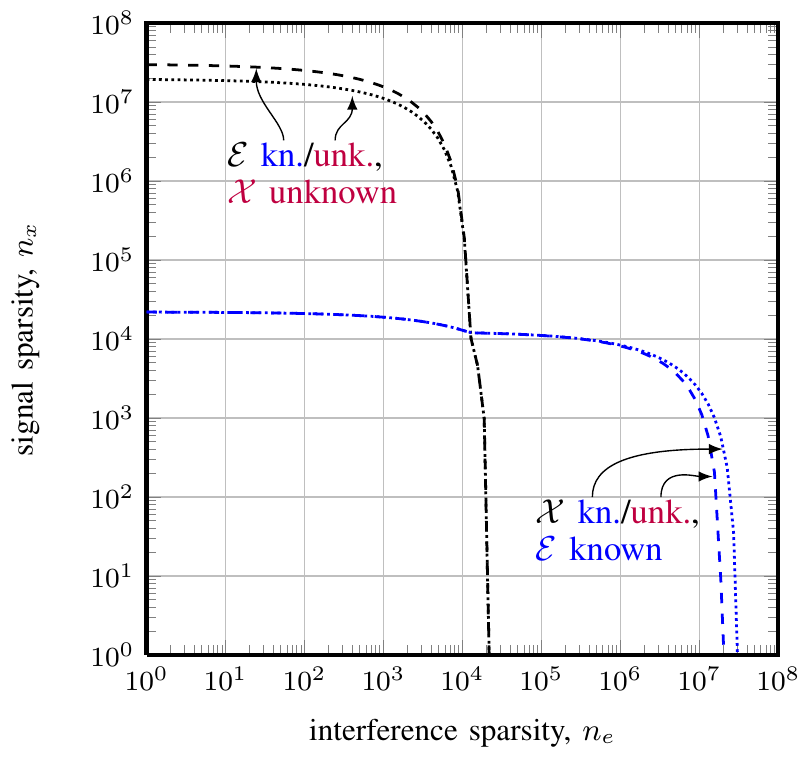}

    \caption{\rev{Impact of support-set knowledge on the recovery conditions for \eqref{eq:both_conc}, \PZ, and \PzeroEs in the case where \sigsup and \errsup are both random. %
In the top left we assume \dicta is unitary and \dictb is the concatenation of two unitary matrices so that $\dimm=\dima=10^8$, $\dimb = 2\dima$, $\coha=0$, and $\cohb=\mcoh = 1/\sqrt{\dimm}$ as described in \cite{Calderbank1997,Gribonval2003a}.
For the curves in the bottom right (with $\sigsup$ known/unknown and \errsup known) we reverse the roles of \dicta and \dictb, so that now \dictb is unitary.}}
    \label{fig:knowledge_etf}
\end{figure}

A similar behavior is also apparent in the probabilistic results presented here. 
Specifically, \rev{for unitary and maximally incoherent \dicta and \dictb}, the recovery conditions in \fref{fig:knowledge}  \rev{using \eqref{eq:both_conc}},  \PZ, and \PzeroEs show a  similar factor-of-two gain in the case where both \sigsup and \errsup are chosen at random.
For example, knowledge of~\sigsup enables one to recover a \rev{pair} $(\sig,\err)$ with approximately twice as many non-zero entries compared to the case of not knowing \sigsup.
\rev{In \fref{fig:knowledge_etf}, we show the recovery conditions for the case 
where one dictionary is unitary, but the other is  a concatenation of two unitary matrices, as described earlier in \fref{sec:discussion}.
We again see that the extra support-set knowledge allows us to guarantee the recovery of a signal with more non-zero entries.
It is interesting to note that in both of these scenarios, by adding the knowledge of one of the support sets, we increase the number of non-zero components we can guarantee to recover in the \emph{other} signal component.
For example, by knowing \sigsup prior to recovery, we can guarantee to recover a signal with more non-zero entries in \err.
}

We note that a similar gain is apparent for~\sigsup arbitrary and \errsup random, as well as  for using  \BP and \PoneEs instead of \PZ and \PzeroEs.
%


\subsection{Asymptotic behavior of the recovery conditions}
\label{sec:asymptconditions}
We now compare the asymptotic behavior of probabilistic and deterministic recovery conditions, i.e., we study the scaling behavior of \nsig and \nerr.
To this end, we are interested in the largest~\nsig for which recovery of \sig (and \err) from \obs can be guaranteed with high probability. In particular, we consider the following models for the sparse interference vector \err:
\begin{inparaenum}[i)]
\item Constant sparsity, i.e., $\nerr = 10^3$,
\item sparsity proportional to the square root of the problem size, i.e., $\nerr = \sqrt{\dimm}$, and 
\item sparsity proportional to the problem size, i.e., $\nerr = \dimm/10^5$.
\end{inparaenum}

\begin{figure}[t]
    \centering
     \includegraphics[height=8cm]{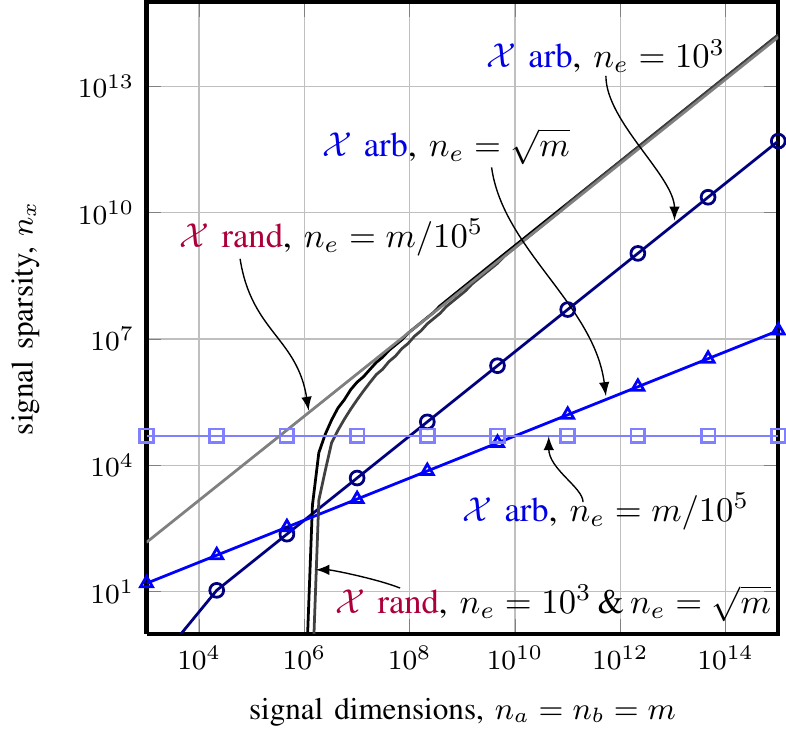}

  \caption{Maximum signal sparsity \nsig that ensures recovery of \vecx for \errsup known and arbitrary. We assume $\nerr = 10^3$, $\nerr = \sqrt{\dimm}$, and $\nerr = \dimm/10^5$. The probability of successful recovery is set to be at least $\rev{1-}10^{-15}$.}
    \label{fig:one_scaling}
\end{figure}

\fref{fig:one_scaling} shows the largest \nsig for which recovery can be guaranteed using \PoneEs. 
Here,~\errsup is assumed to be known and arbitrary and \sigsup is unknown and chosen at random. Note that the other cases of support-set knowledge and arbitrary/random exhibit the same scaling behavior.
We see from \fref{fig:one_scaling} that for a constant interference sparsity (i.e., $\nerr = 10^3$), the probabilistic and deterministic results show the same scaling behavior.
For the cases where \nerr scales with $\sqrt{\dimm}$ or $\dimm$, however, the deterministic thresholds developed in \cite{Studer2012} result in worse scaling, while the behavior of the probabilistic guarantees derived in this paper remain unaffected.

We now investigate the scaling behavior observed in \fref{fig:one_scaling} analytically.
Again, we only consider the case where \sigsup is unknown and chosen at random and \errsup is known and chosen arbitrarily; an analysis of the other cases yields similar results.
\rev{Assume that \dicta and \dictb are unitary and maximally incoherent, i.e., $\coha=\cohb=0$, $\dima=\dimb=\dimm$, and $\mcoh=1/\sqrt{m}$.
Then, by \fref{thm:XurEka_BP}, the recovery of \sig from \obs using \PoneEs  is guaranteed with probability at least $1-3/\dima$  (i.e., for $\beta = \log(\dima)$) if
\begin{align*}
  \delta e^{- 1/4} \ge \sqrt{\nsig/\dima} + 3\mcoh \sqrt{2\beta\nerr},
\end{align*}
 and  
\begin{align*}
  2\nerr\mcoh^2(\log(\dima)+\beta) < (1-\delta)^2,
\end{align*}
hold. 
Combining these two conditions gives
\begin{align} \label{eq:examplecondition1}
e^{-\frac 1 4} \sqrt{\dimm}  > \sqrt{\nsig} + (3\sqrt{2}+2e^{-\frac 1 4})\sqrt{\nerr \log(\dimm)}.
\end{align}}Hence, if $\nsig \sim \dimm$ and $\nerr \sim \dimm/\log(\dimm)$, the condition \fref{eq:examplecondition1} can  be satisfied.
Consequently, recovery of \sig (and of \err) is guaranteed with probability at least $1-3/\dimm$ even if \nsig scales \emph{linearly} in the number of (corrupted) measurements \dimm \emph{and} \nerr scales near-linearly (i.e., with $\dimm/\log(\dimm)$) in~\dimm. 

We finally note that the recovery guarantees in \cite{Li2011} also allow for the sparsity of the interference vector to scale near-linearly in the number of measurements. The results in \cite{Li2011}, however, require the matrix \dicta to be random and \dictb to be orthogonal, whereas the recovery guarantees shown here are for \emph{arbitrary} pairs of dictionaries \dicta and \dictb (characterized by the coherence parameters) and for varying degrees of support-set knowledge.

\subsection{No error component}
\rev{It is worth briefly discussing how our results behave when there is no error, that is when $\nerr=0$.  In this case, the relevant setting is with \sigsup  unknown and chosen uniformly at random.
As \fref{thm:XurEka_P0} holds for any \dictb, it suffices to take \dictb equal to a single column\footnote{Taking \dictb to be the zero-matrix and so removing all the terms that appear in the recovery conditions also leads to the same scaling behavior.}, since $\nerr=0$ means we do not consider any component of \dictb when attempting to recover the signals.
And since the mutual coherence \mcoh only appears as a product with \nerr, it does not matter what we assume \mcoh to be.
Thus by taking $\nerr=0$ and applying \fref{thm:XurEka_P0} we find that for \PzeroEs, recovery is guaranteed with probability at least $1-e^{-\beta}$ if
\begin{align}
e^{-\frac 1 4} (1-\nsig \coha^2) \ge \norm{l2}{\dicta} \sqrt{\frac \nsig\dima} + 12 \coha \sqrt{\beta\nsig}. \label{eq:ne0P0}
\end{align}
For \PoneEs, recovery is guaranteed with probability at least $1-3e^{-\beta}$ if
\begin{align}
e^{-\frac 1 4} \lefto(1-\sqrt{2\nsig \coha^2(\log(\dima)+\beta)}\right) \ge \norm{l2}{\dicta} \sqrt{\frac \nsig\dima} + 12 \coha \sqrt{\beta\nsig}. \label{eq:ne0BP}
\end{align}
Now assume that $\coha\sim 1 /\sqrt{m}$, $\norm{l2}{\dicta}^2 = \dima/\dimm$, and that $\beta = \log(\dima)$.
Then (after ignoring lower order terms), we find that \eqref{eq:ne0P0} and \eqref{eq:ne0BP} imply recovery with probability at least $1-1/\dima$ and $1-3/\dima$, respectively, provided that
\begin{align*}
  \dimm \ge \mathrm{C}\, \nsig \log(\dima),
\end{align*}
for some positive constant $\mathrm{C}$. 
This result is in accordance with \cite{Tropp2008}, the RIP-based proof of \cite{Baraniuk2007b} which requires $\dimm\ge \mathrm{C}_0\,\nsig \log(\dima/\nsig)$ to guarantee recovery with high probability, and the random sub-sampling model of \cite{Candes2007}, which, for a maximally incoherent sparsity basis and measurement matrix\footnote{For example, measuring with a randomly sub-sampled Fourier matrix and taking the Identity matrix as the sparsity basis, so that with the differently normalized definition of coherence as in \cite{Candes2007}, $\coha = 1$.}, requires $  \dimm \ge \mathrm{C}_1\, \nsig \log(\dima)$ to guarantee recovery with high probability.
Thus, our results reduce to some of the existing results in the setting where there is no error.
}


\section{Conclusions}
\label{sec:conc}

In this paper, we have presented novel coherence-based recovery guarantees for sparsely corrupted signals in the probabilistic setting. In particular, we have studied the case  where the sparse signal and/or sparse interference vectors are modeled as random and the dictionaries \dicta and~\dictb are solely characterized by their coherence parameters. 
Our recovery guarantees complete all missing cases of support-set knowledge and improve \rev{and refine}  the results in \cite{Studer2012,Kuppinger2012}.
Furthermore, we have shown that the reconstruction of sparse signals is guaranteed with high probability, even if the number of non-zero entries in both the sparse signal and sparse interference are allowed to scale (near) linearly with the number of (corrupted) measurements.

There are many avenues for follow-on work. The derivation of probabilistic recovery guarantees for the more general setting studied in \cite{Studer2012a}, i.e., $\obs = \dicta\sig+\dictb\err+\vecn$ with $\vecn$ being additive noise and $\vecx$ and $\vece$ being approximately sparse (rather than perfectly sparse), is left for future work.
\rev{In addition, our framework could be generalized to the setting where we split both the known and the unknown support sets into a random and arbitrary part, resulting in four parts, as outlined in \fref{sec:rin}.}
Finally, the derivation of probabilistic uncertainty relations for pairs of general dictionaries is an interesting open problem and would complete the deterministic uncertainty relations in \cite{Kuppinger2012,Studer2012}.

\section*{Acknowledgments}

The authors would like to thank C.~Aubel, R.~G.~Baraniuk, H.~B\"olcskei, I.~Koch, P.~Kuppinger, A.~Pope, and E.~Riegler for inspiring discussions. We would also like to thank the anonymous reviewers for their valuable comments, which improved the overall quality of the paper.




\appendices

\section{Bounds on $\smin(\dictse)$ }
\label{app:smax}

We now derive probabilistic bounds on $\smin(\dictse)$, which are key in showing when the recovery from sparsely corrupted signals succeeds.
We extend  \cite[Lemma 7]{Kuppinger2012} to the case where both supports \sigsup and \errsup are chosen at random and give improved results for the case where only one support set is random.
First, we require the following two results from \cite{Tropp2008}.

\begin{thm}[Thm.~8 of \cite{Tropp2008}] \label{thm:211}
Let $\matM\in\complexset^{\dimm\times \dimn}$ be a matrix.   Let $\support\subseteq\{1,2,\ldots,\dimn\}$ be a set of size $\sparsity$ drawn uniformly at random.  
Fix $q\ge 1$, then for each $p \ge \max\{2,2\log(\rank(\matM\rest_\support)), q/2\}$ we have
\begin{align*}
    \moment{q}{\norm{l2l2}{\matM\rest_\support }}\le 3\sqrt{p} \norm{l1l2}{\matM} + \sqrt{\frac{\sparsity}{\dimn}} \norm{l2l2}{\matM},
\end{align*}
where $\norm{l1l2}{\matM} = \sup_{\vecv\in\complexset^{n}} \norm{l2}{\matM\vecv} / \norm{l1}{\vecv}$ and is the maximum $\ell_2$-norm of the columns of $\matM$.
\end{thm}

\begin{lem}[Eq.~6.1 of \cite{Tropp2008}] \label{lem:momentbound}
Let $\matM\in\complexset^{\dimm\times\dimn}$ be a matrix with coherence $\coh$ and let $\setS\subseteq\{1,2,\ldots,\dimn\}$ be a set of size $s$ chosen uniformly at random. Then, for $\beta \ge \log(s)$ and $q=4\beta$
\begin{align*}
\moment{q}{\norm{l2l2}{\matM_\setS^\adj \matM_\setS - \identity}} \le 12 \coh \sqrt{\beta s} + \isonb \frac{2s}{\dimn}\norm{l2l2}{\matM}^2.
\end{align*}
\end{lem}

\sloppy

Note that the result in \cite[Eq.~6.1]{Tropp2008} does not include the indicator function $\isonb$. It is, however, straightforward to verify that if $\matM$ is orthonormal, then $\mu=0$ and hence, \mbox{$\norm{l2l2}{\matM_\setS^\adj \matM_\setS - \identity} = 0$} for all sets $\setS$.

\fussy

We now state the main result for $\smin(\dictse)$.
\begin{thm} \label{thm:321}
Choose $\beta \ge\log(\nsig)$, $q=4\beta$ and assume that \dicta and \dictb are characterized by the coherence parameters \coha, \cohb, and \mcoh.
If
\begin{inparaenum}[i)]
\item \sigsup is chosen uniformly at random with cardinality~\nsig, \errsup~is arbitrary, and \eqref{eq:smin_ra} holds, or
\item \errsup is chosen uniformly at random with cardinality \nerr,  \sigsup is arbitrary, and \eqref{eq:smin_ar} holds, or
\item both \sigsup and \errsup are chosen uniformly at random with cardinalities \nsig and \nerr respectively, and \eqref{eq:smin_rr} holds,
\end{inparaenum}
then
\begin{align}
    \prob{\norm{l2l2}{\dict_{\sigsup,\errsup}^\adj \dict_{\sigsup,\errsup} - \identity} \ge \delta } \le e^{-\beta},\label{eq:ddbound}
\end{align}
and if \eqref{eq:smin_ra}, \eqref{eq:smin_rr} or \eqref{eq:smin_ar} hold with $\delta=1$, then
\begin{align}
   \prob{\smin(\dictse) =0 } \le e^{-\beta}. \label{eq:smin}
 \end{align}
\end{thm}
\begin{IEEEproof}
The proof follows that of \cite[Lemma 7]{Kuppinger2012}.
We start by defining the hollow Gram matrix 
\begin{align*}
  &\matH = \dictse^\adj \dictse - \identity 
  = \begin{bmatrix} \dicta^\adj_\sigsup\dicta_\sigsup - \identity & \dicta_\sigsup^\adj \dictb_\errsup \\ \dictb^\adj_\errsup\dicta_\sigsup & \dictb^\adj_\errsup \dictb_\errsup - \identity \end{bmatrix}.
\end{align*}
Splitting $\matH$ into diagonal and off-diagonal blocks and applying the triangle inequality leads to
\begin{align*}
\norm{l2l2}{\matH}  & \le \norm{l2l2}{\begin{bmatrix} \dicta^\adj_\sigsup\dicta_\sigsup - \identity & 0\\ 0 & \dictb^\adj_\errsup \dictb_\errsup - \identity \end{bmatrix}} \!\!\!\!\!\! + \norm{l2l2}{\begin{bmatrix} 0 & \dicta_\sigsup^\adj \dictb_\errsup \\ \dictb^\adj_\errsup\dicta_\sigsup & 0\end{bmatrix}} \\
  &\le \max\lefto\{ \norm{l2l2}{\dicta^\adj_\sigsup\dicta_\sigsup - \identity} , \norm{l2l2}{ \dictb^\adj_\errsup \dictb_\errsup - \identity} \right\} + \norm{l2l2}{\dictb^\adj_\errsup\dicta_\sigsup} \\
&\le \norm{l2l2}{\dicta^\adj_\sigsup\dicta_\sigsup - \identity} + \norm{l2l2}{ \dictb^\adj_\errsup \dictb_\errsup - \identity}  + \norm{l2l2}{\dictb^\adj_\errsup\dicta_\sigsup}.
\end{align*}
Since the $q$th moment effectively defines an $\ell_q$-norm, it satisfies the triangle inequality, namely, $\moment{q}{\abs{X+Y}} \le \moment{q}{\abs{X}} + \moment{q}{\abs{Y}}$. Hence, it follows that
\begin{align}
  \moment{q}{\norm{l2l2}{\matH}} \le &\ \moment{q}{\norm{l2l2}{\dicta^\adj_\sigsup\dicta_\sigsup - \identity} }   +\moment{q}{\norm{l2l2}{ \dictb^\adj_\errsup \dictb_\errsup - \identity}} + \moment{q}{\norm{l2l2}{\dictb^\adj_\errsup\dicta_\sigsup}}. \label{eq:one_3terms}
\end{align}
We now separately bound each of the terms in \eqref{eq:one_3terms} and we do this for each case where \sigsup and~\errsup is either chosen at random or arbitrarily.
If \sigsup is chosen uniformly at random, then  it follows from \fref{lem:momentbound} that
\begin{align}
  \moment{q}{\norm{l2l2}{\dicta^\adj_\sigsup\dicta_\sigsup - \identity} } \le 12\coha \sqrt{\beta \nsig} + \isonba\frac{2\nsig}{\dima} \norm{l2l2}{\dicta}, \label{eq:one_firstterm}
\end{align}
for any $4\beta = q \ge 4 \log(\nsig)$.
%
If \sigsup is allowed to be arbitrary, then for all  \sigsup we have
\begin{align}
   \norm{l2l2}{ \dicta^\adj_\sigsup \dicta_\sigsup - \identity} \le \max_k \sum_{j\neq k} \abs{ [\dicta_\sigsup^\adj \dicta_\sigsup]_{j,k}  } \le (\nsig-1) \coha, \label{eq:one_secondterm}
\end{align}
where the first inequality follows from the \Gersgorin disc theorem \cite[Thm.~6.1.1]{Horn1990} and the second inequality is a consequence of the definition of \coha. 
By reversing the role of \dicta and \dictb, we get the analogous bounds for the right-hand side (RHS) term $\moment{q}{\norm{l2l2}{ \dictb^\adj_\errsup \dictb_\errsup - \identity}} $ in \fref{eq:one_3terms}.


 For the third summand appearing in the RHS of \eqref{eq:one_3terms}, let us first consider the case where~$\errsup$ is chosen arbitrarily and \sigsup uniformly at random. 
We then want to apply  \fref{thm:211} to  $\matM = \dictb_\errsup^\adj \dicta$ and $\rest_\sigsup$.
Since $\matM\rest_\sigsup$ \rev{has \nerr rows and \nsig non-zero columns}, $\rank(\matM\rest_\sigsup) \le \min\{\nsig,\nerr\}$ and thus we can apply \fref{thm:211} with $q=2p=4\beta$ where $q \ge  4\min\{\log(\nsig),\log(\nerr)\} \ge 4\log(\rank(\matM\rest_\sigsup)) $ to get
\begin{subequations}
\begin{align}
  \moment{q}{\norm{l2l2}{\dictb^\adj_\errsup\dicta_\sigsup}}& \rev{= \momentx{q}{\sigsup}{\norm{l2l2}{\dictb^\adj_\errsup\dicta_\sigsup}}} \label{eq:xexpect} \\
  &\le  3\sqrt{p} \norm{l1l2}{ \dictb^\adj_\errsup\dicta } + \sqrt{\frac{\nsig}{\dima}} \norm{l2l2}{\dictb_\errsup^\adj \dicta}\nonumber \\
  &\le 3\mcoh\sqrt{2\beta\nerr} + \sqrt{\frac{\nsig}{\dima}} \norm{l2l2}{\dictb^\adj \dicta} ,\label{eq:one_thirdterm}
\end{align}
\end{subequations}
where the entries of $\dictb^\adj_\errsup\dicta$ are bounded by the mutual coherence $\mcoh$.
The case where \errsup is random and \sigsup is arbitrary follows by reversing the roles of \dicta and \dictb.

Now consider the case where both \errsup and \sigsup are random.
\rev{
We can set $\matM = \dictb^\adj \dicta$ so that we may write $\moment{q}{\norm{l2l2}{\dictb^\adj_\errsup\dicta_\sigsup}} =\momentx{q}{\errsup}{\momentx{q}{\sigsup}{\norm{l2l2}{\rest_\errsup \matM \rest_\sigsup} }}$ in order to apply \fref{thm:211}  to first bound the inner expectation, and then to bound the resulting outer expectation.
However, this approach results in a worse bound compared to reusing \eqref{eq:one_thirdterm}, which does not depend on \errsup and hence holds for all $\errsup$. 
By also taking the expectation in \eqref{eq:xexpect} with respect to \errsup instead of \sigsup and bounding similarly, we get that
}
\begin{align}
  \moment{q}{\norm{l2l2}{\dictb^\adj_\errsup\dicta_\sigsup}} \le & \, \min\left\{ 3\mcoh \sqrt{2\beta\nsig} + \sqrt{\frac{\nerr}{\dimb}} \norm{l2l2}{\dicta^\adj \dictb}, \right. \nonumber\\
   &\ \, \qquad \left.3\mcoh \sqrt{2\beta\nerr} + \sqrt{\frac{\nsig}{\dima}} \norm{l2l2}{\dicta^\adj \dictb}\right\}, \label{eq:eab_rr}
\end{align}
for any $\beta \ge \min\{\log(\nsig), \log(\nerr)\}$.
Combining \eqref{eq:one_firstterm}, \eqref{eq:one_secondterm}, \eqref{eq:one_thirdterm}, and \eqref{eq:eab_rr} with the analogous results for \dictb and \errsup leads to the conditions \eqref{eq:smin_ra}, \eqref{eq:smin_rr}, and \eqref{eq:smin_ar}.

Due to \eqref{eq:one_firstterm} and the analogous result for $\dictb_\errsup$, if \sigsup is chosen at random, we require $\beta\ge \log(\nsig)$, if \errsup is chosen at random we need $\beta \ge \log(\nerr)$, and if both \sigsup and \errsup are chosen at random, both of these conditions need to be satisfied, namely that $\beta \ge \max\{\log(\nsig), \log(\nerr)\}$.

We now show that the conditions \eqref{eq:smin_ra}, \eqref{eq:smin_rr}, and \eqref{eq:smin_ar} are sufficient to show that  \eqref{eq:ddbound} holds.
Chebyshev's Inequality \cite[Sec.~1.3]{Durrett2010} states that for a random variable $X$ and a function \mbox{$f\colon\reals \rightarrow \reals^+$}
\begin{align}
    \prob{X \in \setA} \le \frac{\expected{f(X)}}{\inf\lefto\{ f(x) \colon x\in\setA \right\}}. \label{eq:chebyshev}
\end{align}
Application of \eqref{eq:chebyshev} with $f(x) = x^q$ and the random variable $X =\norm{l2l2}{\dict_\support^\adj \dict_\support - \identity}$ gives
\begin{align}
    \prob{X \ge \delta } &\le \frac{\expected{X^q}}{ \inf\lefto\{ x^q \colon x\ge \delta \right\}} 
\le \frac{\left(\delta e^{-1/4}\right)^q}{\delta^q}  = e^{-q/4}, \label{eq:qbound}
\end{align}
provided that $(\delta e^{-1/4})^q \ge \expected{X^q}$. 
But this is guaranteed by the assumptions in \eqref{eq:smin_ra},  \eqref{eq:smin_rr}, or \eqref{eq:smin_ar}, depending on the  signal and interference model.
Therefore, we have
\begin{align*}
  \prob{\norm{l2l2}{\matH} \ge \delta } \le e^{-\beta},
\end{align*}
since $q=4\beta$.
The second part of the theorem, \eqref{eq:smin}, is a result of the fact that $\smin(\dictse) = 0$ implies that $\norm{l2l2}{\matH} \ge 1$ and hence, $ \prob{\smin(\dictse)=0}\le  \prob{\norm{l2l2}{\matH} \ge 1 } $.
\end{IEEEproof}

\section{Both Supports Known}
\label{app:XEknown}


\begin{IEEEproof}[Proof of  \fref{thm:XkrEka}]
It suffices to show that $\dictse$ is invertible, which is equivalent to the condition that $\smin(\dictse)>0$.
By assumption, the conditions of \fref{thm:321} hold, which implies $\prob{\smin(\dictse) = 0} \le e^{-\beta}$. Hence, recovery of \sig and \err using \eqref{eq:both_conc} succeeds with probability at least $1-e^{-\beta}$.
\end{IEEEproof}

\section{\PZ with Limited Support Knowledge}
\label{app:PZ}
We now prove the recovery guarantees for \PZC, \PzeroEs, and \PzeroX for partial (or no) support-set knowledge of \errsup and \sigsup.
We follow the proof of \cite{Tropp2008} and present the three cases
\begin{inparaenum}[1)]
\item \sigsup known,
\item \errsup known, and
\item no support-set knowledge, 
\end{inparaenum}
all together, since the corresponding proofs are similar.
Note that $\setR(\dict)$ denotes the space spanned by the columns of~\dict.

We begin by generalizing \cite[Thm.~13]{Tropp2008} to the case of pairs of dictionaries \dicta and \dictb where we know the support set of \err.
The result gives us a sufficient condition for when there is a unique minimizer of \PZC, \PzeroEs, or \PzeroX.
\begin{lem}[Based on Thm.~13 of \cite{Tropp2008}]\label{lem:P0_uniq}
Let $\dictaa\in\complexset^{\dimm\times\dima}$ and $\dictbb\in\complexset^{\dimm\times\dimb}$ be two dictionaries and suppose that we observe the signal $\obs = \dictaa \sig + \dictbb \err$ where $\sigsup=\supp(\sig)$ and $\errsup = \supp(\err)$ and the non-zero entries of \sig \rev{and \err} are drawn from a continuous distribution.
Furthermore, suppose that \errsup is known.
\rev{Write $\dictt=[\,\dictaa\,\,\dictbb\,]$ and $\dictset=[\,\dictaa_\sigsup\,\,\dictbb_\errsup\,]$.}
%
%
%
If
\begin{align}
    \dim\lefto( \range\lefto(\dictset\right) \cap \range\lefto(\tilde{\dict}_{\sigsup',\errsup}\right) \right) < \abs{\sigsup} + \abs{\errsup}, \label{eq:range_bound}
\end{align}
for all sets \rev{$\sigsup'\neq \sigsup$} where $\abs{\sigsup} = \abs{\sigsup'}$, then, almost surely, \PzeroEs recovers  the vectors \sig and~\err.
\end{lem}
This result also provides a sufficient condition for \PZC, if we set $\dictaa=\dict$ and take \dictbb to be the empty matrix, or for \PzeroX, if we set $\dictaa=\dictb$ and $\dictbb=\dicta$.

\begin{IEEEproof}
We follow the proof of \cite[Thm.~13]{Tropp2008}. 
\rev{We begin by defining the set of all  alternative representations as follows:
 \begin{align*}
   \setD_{\sigsup,\sigsup'}^{\errsup} \define \lefto\{ (\sig,\err) \colon \begin{array}{l}\dictaa\sig+\dictbb\err = \dictaa\sig'+\dictbb\err'\\ \supp(\sig) = \sigsup,\ \supp(\sig') = \sigsup'\\ \supp(\err) = \supp(\err') = \errsup \end{array}\right\},
 \end{align*}
 and the set of observations that have alternative representations 
 \begin{align*}
     \setAss \define \left\{ \obs \colon \obs = \dictaa_\sigsup\sig_\sigsup + \dictbb_\errsup\err_\errsup, (\sig,\err)\in\setD_{\sigsup,\sigsup'}^{\errsup}   \right\},  
 \end{align*} 
}
so that $\setAss$ is the set of observations that can be written in terms of two pairs of signals $(\sig,\err)$ and $(\sig', \err')$ where $\sigsup=\supp(\sig)$, $\sigsup'=\supp(\sig')$, and $\errsup=\supp(\err)=\supp(\err')$.

For any $\sigsup'$ of size $\abs{\sigsup}$ and $\sigsup'\neq\sigsup$, we have
\begin{align*}
  \setAss \subseteq \range\lefto(\dictset\right) \cap \range\lefto(\tilde{\dict}_{\sigsup',\errsup}\right).
\end{align*}
Now assume that  \eqref{eq:range_bound} holds for $\sigsup$, $\sigsup'$, and $\errsup$,  then $\dim(\setAss) < \abs{\sigsup} + \abs{\errsup}$.
Thus the smallest subspace containing $\setAss$ is a strict subspace of $\range(\dictset)$ and hence, has zero measure with respect to any nonatomic measure \rev{defined in the range of $\dictset$}. 
Since $\sig$ \rev{and \err}, and hence \obs, have non-zero entries drawn from a continuous distribution
\begin{align*}
  \prob{\dictaa\sig+\dictbb\err =\obs \in \setAss}=0.
\end{align*}
Thus, with probability zero, there exists no alternative pair $(\sig',\err')$ with supports $\sigsup'$ and \errsup, respectively, otherwise $\obs$ would lie in $\setAss$.
\rev{Therefore, if} \eqref{eq:range_bound} holds for all $\sigsup'$, then \rev{the probability of choosing random \sig and \err so that \obs admits an alternative representation is zero, and hence}, almost surely, given $\obs=\dictaa\sig+\dictbb\err$, \PzeroEs returns the vectors \sig and \err.
\end{IEEEproof}

We can use \fref{lem:P0_uniq} to prove the first part of Theorems \ref{thm:XurEka_P0}, \ref{thm:XurEkr_P0}, \ref{thm:XkrEua_P0}, \ref{thm:XurEua_P0}, and \ref{thm:XurEur_P0} by showing that~\fref{eq:range_bound} holds with high probability.
To show that \eqref{eq:range_bound} holds for all $\sigsup'$ we show that for every column $\tilde{\cola}_\gamma$ of $\dictaa$ not in $\dictaa_\sigsup$ (i.e., for all $\gamma\notin\sigsup$) that $\tilde{\cola}_\gamma \notin \range(\dictset)$, which is equivalent to showing that
\begin{align}
    \norm{l2}{\projset \tilde{\cola}_\gamma} < \norm{l2}{\tilde{\cola}_\gamma} = 1, \label{eq:dgamma}
\end{align}
for all $\gamma\notin\sigsup$ and where $\projset=(\pinv{\dictset})^{\rev{\adj}} \dictset^\adj$ 
is the projection onto the range space of $\dictset$.
We will now bound the probability that \eqref{eq:dgamma} holds for the following three situations: 
\begin{inparaenum}[1)]
\item only \errsup known, 
\item only \sigsup known, and 
\item both support sets unknown. 
\end{inparaenum}

\subsubsection{Only \errsup known}
Consider the setting where \errsup is known, but \sigsup is unknown; this case fits the setting of \fref{lem:P0_uniq} with $\dictaa=\dicta$ and $\dictbb=\dictb$.
Hence, the condition \eqref{eq:dgamma} is equivalent to $\norm{l2}{\projse \cola_\gamma} < \norm{l2}{\cola_\gamma} =1$.
We have
\begin{align*}
    \norm{l2}{\projse \cola_\gamma} &\le \norm{l2l2}{(\pinv{\dictse})^\adj} \norm{l2}{\dictse^\adj \cola_\gamma} \\
    & \le \smin^{-1}(\dictse) \sqrt{\norm{l2}{\dicta_\sigsup^\adj \cola_\gamma}^2+\norm{l2}{\dictb_\errsup^\adj \cola_\gamma}^2}.
\end{align*}
From the definitions of the coherence parameters\footnote{Note that we use bounds that hold for \emph{all} \sigsup, rather than a bound that holds with high probability. The underlying reason is the fact that if \dicta is an equiangular tight frame, the associated inequalities hold with equality and hence, we cannot do any better by using probabilistic bounds, unless we take advantage of a property of \dicta other than the coherence \coha.} 
\begin{align}
    \norm{l2}{\dictse^\adj \cola_\gamma} \le \xi_\errsup \define \sqrt{\coha^2 \nsig + \mcoh^2 \nerr}. \label{eq:one_errsup}
\end{align}
Thus, in order to guarantee $\norm{l2}{\projse \cola_\gamma} < 1$ it suffices to have 
\begin{align}
 \xi_\errsup < \smin(\dictse). \label{eq:P0_Eknown}
\end{align}

\subsubsection{Only \sigsup known}
For the setting where only \sigsup is known, we apply \fref{lem:P0_uniq} with $\dictaa=\dictb$ and $\dictbb=\dicta$, thus the condition of \eqref{eq:range_bound} becomes $
  \dim\lefto( \range\lefto(\dictse\right) \cap \range\lefto({\dict}_{\sigsup,\errsup'}\right) \right) < \abs{\sigsup} + \abs{\errsup}$,
and so we only want to show that $\norm{l2}{\projse \colb_\gamma} < \norm{l2}{\colb_\gamma} $ for all $\gamma\notin\errsup$.
Proceeding as before, it follows that
\begin{align}
    \norm{l2}{\projse \colb_\gamma} 
    & \le \smin^{-1}(\dictse) \norm{l2}{\dictse^\adj \colb_\gamma} \nonumber\\
    &\le \smin^{-1}(\dictse)\, \xi_\sigsup, \label{eq:one_sigsup}
\end{align}
where $\xi_\sigsup \define \sqrt{\mcoh^2\nsig + \cohb^2\nerr}$. Hence, it suffices to show that
\begin{align}
 \xi_\sigsup < \smin(\dictse). \label{eq:P0_Xknown}
\end{align}

\subsubsection{No support-set knowledge}
Finally, we consider the setting where neither \sigsup nor \errsup is known, so we apply \fref{lem:P0_uniq} with $\dictaa = [\,\dicta\,\, \dictb\,]$ and $\dictbb$ being the empty matrix, thus this is exactly the condition of \cite[Thm.~13]{Tropp2008}.
Then, we show that  $\norm{l2}{\projse \dcol_\gamma} < \norm{l2}{\dcol_\gamma} $ for any column~$\dcol_\gamma$ of \dict not in $\dictse$.
In other words, we want both \eqref{eq:one_errsup} and \eqref{eq:one_sigsup} to hold as $\dcol_\gamma$ can be a column of either \dicta or \dictb.
So it suffices to show
\begin{align}
  \norm{l2}{\projse \dcol_\gamma} \le \smin^{-1}(\dictse)\, \xi_{+} < 1, \label{eq:P0_noneknown}
\end{align}
where $\xi_{+} = \max\{\xi_\sigsup,\xi_\errsup\}$.

Finally, to show that the \PZ based problems succeed, we want to bound the probability that \eqref{eq:P0_Eknown}, \eqref{eq:P0_Xknown}, or \eqref{eq:P0_noneknown} holds (depending on which, if any, support sets we know).
In each of the cases, we know that \PZC, \PzeroEs, or \PzeroX returns the correct solution if \mbox{$\xi < \smin(\dictse)$}, where $\xi\in(0,1)$ is  equal to $\xi_\errsup$, $\xi_\sigsup$, or $\xi_{+}$ (as appropriate to the case).
Hence, we can bound the probability of error as follows
\begin{align*}
  \prob{\text{error}} & \le \prob{\xi \ge \smin(\dictse)} \\
 &\le \prob{\norm{l2l2}{\dictse^\adj \dictse - \identity}\ge 1-\xi^2 } \le e^{-\beta}, 
\end{align*}
where we use \fref{thm:321} with $\delta = 1-\xi^2$. Therefore, with probability exceeding $1-e^{-\beta}$, the pair $(\sig,\err)$ is the unique minimizer.

\section{\BP with Limited Support Knowledge}
\label{app:bp}
\setcounter{subsubsection}{0}
We now prove the recovery results for the \BP based algorithms.
\rev{To do this, we restate the sufficient recovery condition of \cite{Tropp2005} and then show when we can satisfy this condition, thereby guaranteeing the successful recovery of \sig with \PoneEs, \PoneX, or \BPC.}
\begin{thm}[Thm.~5 of \cite{Tropp2005}] \label{thm:bp_suff}
\rev{Suppose that the sparsest representation of a complex vector \obs is $\tilde\dict_\setS\vecs_\setS$.
If $\dictse$ is full rank and there exists a vector $\vech\in \complexset^\dimm$ such that
\begin{subequations}
\begin{align}
&\tilde\dict_\setS^\adj \vech = \sign(\vecs_{\setS}), \, \text{and} \label{eq:tropp1}\\
&|\langle{\vech},{\tilde\dcol_\gamma}\rangle|< 1 \text{ for all columns } \tilde\dcol_\gamma \text{ of } \dictt \text{ not in } \tilde\dict_\setS,  \label{eq:tropp2}
\end{align}
\end{subequations}
then $\vecs$ is the unique minimizer of \BP.
}
\end{thm}

\rev{We can easily apply \fref{thm:bp_suff} to attain recovery conditions for \PoneEs, \PoneX, and \BPC.
For \PoneEs, we apply \fref{thm:bp_suff} to the matrix $\tilde{\dict} = [\,\dicta\,\,\dictb_\errsup\,]$ so that the two problems \BP and \PoneEs are the same. 
We want to show that $\vecs_\setS^\trans = [\,\sig_\sigsup^\trans\,\,\err_\errsup^\trans\,]$ is the sparsest representation of the observation \obs.
By rewriting \eqref{eq:tropp1} and \eqref{eq:tropp2} it follows that it is sufficient to guarantee recovery with \PoneEs if there exists a vector  $\vech\in \complexset^\dimm$ such that
\begin{subequations}
\begin{align}
&[\,\dicta_\sigsup\,\,\dictb_\errsup\,]^\adj \vech = \sign\lefto(\begin{bmatrix}\sig_\sigsup \\ \err_\errsup \end{bmatrix}\right), \, \text{and} \label{eq:PoneE1}\\
&|\langle{\vech},{\cola_\gamma}\rangle|< 1 \text{ for all columns } \cola_\gamma \text{ of } \dicta \text{ not in } \dicta_\sigsup.  \label{eq:PoneE2}
\end{align}
\end{subequations}
Similarly, to get a recovery condition for \BPC, we merely apply \fref{thm:bp_suff} to the matrix $\tilde\dict = [\,\dicta\,\,\dictb\,]$.
}

Finally, before we can prove the  probabilistic recovery guarantees for the $\ell_1$-norm-based algorithms of Theorems \ref{thm:XurEka_P0}, \ref{thm:XurEkr_P0}, \ref{thm:XkrEua_P0}, \ref{thm:XurEua_P0}, and \ref{thm:XurEur_P0}, we require the following lemma.
\begin{lem}[Bernstein's Inequality, Prop.~16 of \cite{Tropp2008}] \label{lem:bernstein}
Let $\vecv\in\complexset^n$ and let $\boldsymbol{\eps}\in\complexset^n$ be a Steinhaus sequence.  Then, for  $u\ge 0$ we have
\begin{align}
    \prob{\abs{\sum_{i=1}^n \eps_i v_i}\ge u \norm{l2}{\vecv}} \le 2 \exp\lefto({- \frac{u^2}{2}}\right).
\end{align}
\end{lem}

A Steinhaus sequence is a (countable) collection of independent complex-valued random variables, whose entries are  uniformly distributed on the unit circle~\cite{Tropp2008}.

We now prove the second part of Theorems \ref{thm:XurEka_P0}, \ref{thm:XurEkr_P0}, \ref{thm:XkrEua_P0}, \ref{thm:XurEua_P0}, and \ref{thm:XurEur_P0}.
To show that recovery with \BPC, \PoneEs, or \PoneX succeeds, we demonstrate that the vector $\vech$, as in \rev{\fref{thm:bp_suff}}, exists with high probability.
We now consider the following three settings in turn:
\begin{inparaenum}[1)]
\item  only \errsup known, 
\item  only \sigsup known, and 
\item both support sets  unknown. 
\end{inparaenum}
But first, let us assume that in each case~$\dictse$ is full rank.

\subsubsection{Only \errsup known}
Consider the case where \errsup is known but \sigsup is unknown, we show that a vector \vech exists that satisfies \rev{\eqref{eq:PoneE1} and \eqref{eq:PoneE2}} with high probability.
To this end, set~$\vech = \dictse\left(\dictse^\adj \dictse\right)^{-1}\sign(\ssigse)$, so that \eqref{eq:PoneE1} is satisfied. 
Then, for any column $\cola_\gamma$ of~\dicta where $\gamma\notin\sigsup$,
\begin{align*}
    \abs{\scalprod{}{\vech}{\cola_\gamma} }&= \abs{\scalprod{}{\rev{\dictse\left(\dictse^\adj \dictse\right)^{-1}}\sign(\ssigse)}{\cola_\gamma}} \\
    &= \abs{\scalprod{}{\sign(\ssigse)}{\left(\dictse^\adj \dictse\right)^{-1}\dictse^\adj \cola_\gamma} } = \abs{\sum_{j=1}^{\nsig} \eps_j v^\gamma_j },
\end{align*}
with $\boldsymbol{\eps} = \sign(\ssigse)$ and $\vecv^\gamma = \left(\dictse^\adj \dictse\right)^{-1}\dictse^\adj \cola_\gamma$.
%
%
%
Since $\boldsymbol{\eps}$ is a Steinhaus sequence (by assumption), we can apply \fref{lem:bernstein} with  $u = \norm{l2}{\vecv^\gamma}^{-1}$ to arrive at
\begin{align}
    \prob{\abs{\sum_{j=1}^{\nsig} \eps_j v^\gamma_j } \ge 1 } \le 2 \exp\lefto(-\frac{1}{2  \norm{l2}{\vecv^\gamma}^{2}}\right). \label{eq:cb_v}
\end{align}
But we have that
\begin{align*}
    \norm{l2}{\vecv^\gamma}^2 &= \norm{l2}{\left(\dictse^\adj \dictse\right)^{-1}\dictse^\adj \cola_\gamma}^2 \\
    &\le \norm{l2l2}{\left(\dictse^\adj \dictse\right)^{-1}}^2 \norm{l2}{\dictse^\adj \cola_\gamma}^2\le \smin^{-4} (\dictse) \, \xi^2_\errsup,
\end{align*}
where $\xi^2_\errsup = \nsig \coha^2 + \nerr\mcoh^2$.
Hence, \eqref{eq:cb_v} results in
\begin{align*}
    \prob{\abs{\sum_{j=1}^{\nsig} \eps_j v^\gamma_j } \ge 1 } &\le 2 \exp\lefto( - \frac{\smin^4(\dictse) }{2 \xi_\errsup^2 }\right). 
\end{align*}
Now we want \eqref{eq:tropp2} to hold for all $\gamma \notin \sigsup$. Hence, applying the union bound to the result above leads to
\begin{align}
    \prob{\max_{\gamma\notin \sigsup} \abs{\sum_{j=1}^{\nsig} \eps_j v^\gamma_j } \ge 1 } \le 2\dima \exp\lefto( - \frac{\smin^4(\dictse) }{2 \xi_\errsup^2 }\right). \label{eq:BP_Eknown}
\end{align}
 
\subsubsection{Only \sigsup known}
Consider the setting where \sigsup is known, but \errsup is unknown.  \rev{This setting follows exactly as in the setting where \errsup is known and \sigsup is unknown by switching the roles of \sigsup and \errsup.
Thus, we arrive at
\begin{align}
    \prob{\max_{\gamma\notin \errsup} \abs{\sum_{j=1}^{\nerr} \eps_j v^\gamma_j } \ge 1 } \le 2\dimb \exp\lefto( - \frac{\smin^4(\dictse) }{2 \xi_\sigsup^2 }\right), \label{eq:BP_Xknown}
\end{align}
where $\xi^2_\sigsup = \nsig\mcoh^2 + \nerr \cohb^2$ and $\vecv^\gamma = \pinv{\dictse}\colb_\gamma$.
}

\subsubsection{No support-set knowledge}
Finally, we consider the third setting where neither \sigsup nor \errsup are known. In particular, we want to show that in \fref{thm:bp_suff}, we can satisfy \eqref{eq:tropp1} and \eqref{eq:tropp2} with high probability.
For any column $\dcol_\gamma$ of \dict not in $\dictse$, set $\vecv^\gamma = \pinv{\dictse} \dcol_\gamma$. 
In this case, we have
\begin{align*}
    \norm{l2}{\vecv^\gamma}^2 &\le \norm{l2l2}{\left(\dictse^\adj \dictse\right)^{-1}}^2 \norm{l2}{\dictse^\adj \dcol_\gamma}^2 \\
    &\le \smin^{-4} (\dictse) \,{\xi_{+}^2},
\end{align*}
where $\xi^2_+ = \max\{\nsig\coha^2+\nerr\mcoh^2, \nsig\mcoh^2 + \nerr\cohb^2\}$ and hence, 
\begin{align*}
    \prob{\abs{\sum_{j=1}^{\nsig+\nerr} \eps_j v^\gamma_j } \ge 1 } 
    &\le 2 \exp\lefto( - \frac{\smin^4(\dictse) }{2 \xi_{+}^2 }\right). 
\end{align*}
Finally, we want \eqref{eq:tropp2} to hold for all $\dcol_\gamma$. Therefore, applying the union bound to the result above leads to
\begin{align}
    &\prob{\max_{\gamma\notin \sigsup\cup\errsup} \abs{\sum_{j=1}^{\nsig+\nerr} \eps_j v^\gamma_j } \ge 1 }
    \le 2(\dima+\dimb) \exp\lefto( - \frac{\smin^4(\dictse) }{2 \xi_{+}^2 }\right). \label{eq:BP_unknown}
\end{align}


We now want to derive an upper bound on the right hand sides of  \eqref{eq:BP_Eknown}, \eqref{eq:BP_Xknown}, and \eqref{eq:BP_unknown}.
%
%
First we calculate the  probability conditioned on \mbox{$\smin(\dictse) > \lambda\in (0,1)$}.
Note that if $\lambda>0$, then $\smin(\dictse)>\lambda>0$ and we satisfy the remaining assumption of \fref{thm:bp_suff}, namely that $\dictse$ is full rank.

For convenience, in the case where \errsup is known, let us set $N=\dima$ and $\xi = \xi_\errsup$.
In the case where $\sigsup$ is known, set $N=\dimb$ and $\xi = \xi_\sigsup$  and finally, in the case where neither \sigsup nor \errsup are known, set $N =\dima+\dimb$ and $\xi=\xi_{+}$.

Thus, we have
\begin{align}
    &\prob{\max_{\gamma\notin \setS} \abs{\sum_{j=1}^{N} \eps_j v^\gamma_j } \ge 1 \Big \vert \smin(\dictse)>\lambda } 
    \le 2 N  \exp\lefto( - \frac{\lambda^4}{2 \xi^2 }\right) \le 2 e^{-\beta}, \label{eq:cb_v_bound}
\end{align}
for some $\beta \le \lambda^4/({2 \xi^2 }) - \log N$. 

For our particular choice of \vech, \eqref{eq:PoneE1} (in the case where \sigsup or \errsup is known) or \eqref{eq:tropp1} (in the case where both supports are unknown) will always be satisfied.
So let $\event$ be the event that \eqref{eq:PoneE2} (in the case where one support is unknown) or \eqref{eq:tropp2} (in the case where both supports are known) is not fulfilled with our choice of $\vech$ and let $\fr$ be the event that $\dictse$ is not full rank.
As $\event\cup\fr$ is a necessary condition for the \BP based algorithms not to be able to recover the vectors \sig and \err, $\prob{\event\cup\fr}$ is an upper bound on the probability of error.
Then, since $\smin(\dictse)>\lambda>0$ implies that $\fr$ cannot occur, and hence that $\prob{\event\cup\fr \big\vert \smin(\dictse)>\lambda} = \prob{\event\big\vert \smin(\dictse)>\lambda}$, we have that for any $\lambda>0$ 
\begin{align}
    \prob{\event\cup\fr} =&\ \prob{\event\cup\fr \big\vert \smin(\dictse)>\lambda} \prob{\smin(\dictse)>\lambda}  \nonumber\\
    & + \prob{\event\cup\fr \big\vert \smin(\dictse)\le \lambda}  \prob{\smin(\dictse)\le \lambda} \nonumber \\
   \le&\ \prob{\event \big\vert \smin(\dictse)>\lambda} +  \prob{\smin(\dictse)\le \lambda}. \label{eq:erbound}
  \end{align}
We can bound the first summand in \eqref{eq:erbound} using \eqref{eq:cb_v_bound} under the assumption that $\beta\le \lambda^4/({2 \xi^2 }) - \log N$.
The second term we can bound using \fref{thm:321} with $\delta=1-\lambda^2\in(0,1)$, which, provided that $\beta \ge N'$ where $N'$ is the size of the supports chosen at random, says that $  \prob{\smin(\dictse)\le \lambda} \le e^{-\beta}$.
Therefore, we have
\begin{align}
    \prob{\event\cup\fr}  \le 3e^{-\beta},    \label{eq:one_threefac}
\end{align}
and hence, we can recover \sig and \err with probability at least $1-3e^{-\beta}$.

\bibliographystyle{IEEEtran} 

\bibliography{psr}

\end{document}